\documentclass[showpacs,amsmath,amssymb,aps,pra,
longbibliography,
10pt,reprint,superscriptaddress]{revtex4-1}
\usepackage{bm}
\usepackage[breaklinks=true,colorlinks=true,linkcolor=blue,urlcolor=blue,citecolor=blue]{hyperref}
\usepackage{dcolumn}
\usepackage{amsmath,amssymb}
\usepackage{mathdots}
\usepackage{natbib}
\usepackage{soul,color}
\usepackage{graphicx}
\usepackage{epstopdf}
\usepackage[note-name]{notes2bib}
\usepackage{mathptmx}

\begin{document}

\title{Rising speed limits for fluxons via edge quality improvement in wide MoSi thin films}

\author{B.~Budinska}
    \email[Corresponding author:]{barbora.budinska@univie.ac.at}
    \affiliation{Faculty of Physics, University of Vienna, 1090 Vienna, Austria}
\author{B.~Aichner}
    \affiliation{Faculty of Physics, University of Vienna, 1090 Vienna, Austria}
\author{D.\,Yu. Vodolazov}
    \affiliation{Institute for Physics of Microstructures, Russian Academy of Sciences, Nizhny Novgorod region 603087, Russia}
\author{M.\,Yu. Mikhailov}
    \affiliation{B. Verkin Institute for Low Temperature Physics and Engineering,
National Academy of Sciences of Ukraine, 61103 Kharkiv, Ukraine}
\author{F.~Porrati}
    \affiliation{Physikalisches Institut, Goethe University, 60438 Frankfurt am Main, Germany}
\author{M.~Huth}
    \affiliation{Physikalisches Institut, Goethe University, 60438 Frankfurt am Main, Germany}
\author{A.\,V.~Chumak}
    \affiliation{Faculty of Physics, University of Vienna, 1090 Vienna, Austria}
\author{W.~Lang}
    \affiliation{Faculty of Physics, University of Vienna, 1090 Vienna, Austria}
\author{O.\,V.~Dobrovolskiy}
    \email[Corresponding author:]{oleksandr.dobrovolskiy@univie.ac.at}
    \affiliation{Faculty of Physics, University of Vienna, 1090 Vienna, Austria}
\date{\today}

\begin{abstract}
Ultra-fast vortex motion has recently become a subject of extensive investigations, triggered by the fundamental question regarding the ultimate speed limits for magnetic flux quanta and enhancements of single-photon detectors. In this regard, the current-biased quench of a dynamic flux-flow regime -- flux-flow instability (FFI) -- has turned into a widely used method for the extraction of information about the relaxation of quasiparticles (unpaired electrons) in the superconductor. However, the large relaxation times $\tau_\epsilon$ deduced from FFI for many superconductors are often inconsistent with the fast relaxation processes implied by their single-photon counting capability. Here, we investigate FFI in $15$\,nm-thick $182\,\mu$m-wide MoSi strips with rough and smooth edges produced by laser etching and milling by a focused ion beam. For the strip with smooth edges we deduce, from the current-voltage ($I$-$V$) curve measurements, a factor of 3 larger critical currents $I_\mathrm{c}$, a factor of 20 higher maximal vortex velocities of 20\,km/s, and a factor of 40 shorter $\tau_\epsilon$. We argue that for the deduction of the intrinsic $\tau_\epsilon$ of the material from the $I$-$V$ curves, utmost care should be taken regarding the edge and sample quality and such a deduction is justified only if the field dependence of $I_\mathrm{c}$ points to the dominating edge pinning of vortices.
\end{abstract}
\maketitle

\maketitle

\section{Introduction}
Ultra-fast vortex motion in superconductors has recently attracted great attention both experimentally\,\cite{Emb17nac,Rou18mat,Dob19pra,Leo20sst,Dob20nac,Ust20sst} and theoretically\,\cite{Vod19sst,Bez19prb,Kog20prb,Pat20prb,Kog21prb,Pat21prb}. This attention is triggered by the fundamental question regarding the ultimate speed limit for magnetic flux transport via superconducting vortices and the phenomena of generation of sound \cite{Ivl99prb,Bul05prb} and spin waves \cite{Bes14prb,Dob21arx} by fluxons moving at velocities of a few km/s. Furthermore, high-velocity vortex dynamics is essential for the photoresponse
of superconductor microstrip single-photon detectors (SMSPDs). Operated at large bias currents\,\cite{Kor18pra,Kor20pra,Cha20apl,Chi20apl}, which enable a vortex-assisted mechanism of single-photon counting\,\cite{Vod17pra}, such microstrips appear as viable candidates for various applications requiring large-area detectors, e.g. free-space quantum cryptography, deep space optical communication, etc. If an SMSPD is capable of carrying a close-to-depairing transport current, its intrinsic detection efficiency is predicted to reach almost 100\%\,\cite{Vod17pra}. Thus, materials with large critical currents are in demand for SMSPDs and require the understanding of vortex matter under far-from-equilibrium conditions. In this regard, current-biased quenches of the dynamic flux-flow regime provide a way for the extraction of information about the relaxation of quasiparticles (unpaired electrons)\,\cite{Leo20sst} in the superconductor. The deduced quasiparticle relaxation times are then often used for judging whether a material could be potentially suitable for the use in single-photon detectors\,\cite{Cap17apl,Dob20nac,Hof21tsf,Liu21sst,Cir21prm}.

The physical phenomenon underlying this judgement is known as the flux-flow instability (FFI)\,\cite{Emb17nac,Rou18mat,Dob19pra,Leo20sst,Dob20nac,Ust20sst,Hof21tsf,Liu21sst,Vod19sst,Bez19prb,Kog20prb,Pat20prb,Kog21prb,Pat21prb,Cap17apl,Cir21prm}.
FFI occurs in the regime of fast vortex motion and, in the current-driven regime, it becomes apparent as a sudden jump to a highly-resistive state in the current-voltage ($I$-$V$) curve of the superconductor. Microscopically, FFI is associated with the diffusion of quasiparticles from the vortex cores to the surrounding superconducting medium, and the subsequent retrapping of the quasiparticles by other vortices. According to Larkin and Ovchinnikov (LO)\,\cite{Lar75etp,Lar86inb,Bez92pcs}, close to the superconducting transition temperature $T_\mathrm{c}$, the maximal vortex velocity (instability velocity) $v^\ast$ is linked with the quasiparticle energy relaxation time $\tau_\epsilon$ via the expression $v^\ast= [D(14\zeta(3))^{1/2}(1-T/T_\mathrm{c})^{1/2}/(\pi \tau_\epsilon)]^{1/2}$, where $D$ is the electron  diffusion coefficient and $\zeta(x)$ is the Riemann zeta function. This LO expression suggests that materials with shorter $\tau_\epsilon$ should, in principle, allow for higher vortex velocities. The essential assumptions in the LO theory\,\cite{Lar75etp,Lar86inb} were (i) the perfect periodicity of the vortex lattice, (ii) the absence of defects and edges in the superconductor, and (iii) neglect of heating effects associated with the Joule dissipation. Accordingly, the LO theory can take into account neither collective effects leading to the dynamic transformation of the vortex lattice nor vortex-pinning and edge-barrier effects. In subsequent studies, however, it was revealed that larger $v^\ast$ can be realized in materials with a weak uncorrelated
disorder\,\cite{Sil12njp,Shk17prb,Dob17sst}, via the vortex guiding effect\,\cite{Dob19pra} or via addition of a GHz-frequency ac current to the dc bias current\,\cite{Dob20cph}. At the same time, in a uniform system, the highest vortex velocities were experimentally observed in a direct-write Nb-C superconductor with a small diffusion coefficient and fast relaxation of quasiparticles ($\tau_\varepsilon\simeq 20$\,ps)\,\cite{Dob20nac} and in bridges of lead\,\cite{Emb17nac} which has a large diffusion coefficient and a short inelastic electron-phonon relaxation time ($\tau_\mathrm{ep}\simeq 20$\,ps)\,\cite{Wat81ltp}. Remarkably, while many dirty superconductors
(NbN\,\cite{Kor20nph}, MoSi\,\cite{Kor20pra}, NbRe\,\cite{Cir20apl}, NbReN\,\cite{Cir21prm} etc.) possess good single-photon detection capability, $v^\ast$ and $\tau_\epsilon$ deduced from FFI are not always consistent with the fast relaxation processes implied by photon-counting experiments. For instance, the deduced $v^\ast$ values of $\simeq 1$\,m/s for $\alpha$-MoSi\,\cite{Liu21sst,Sam95prl}, $0.1$-$0.5$\,km/s for MoSi\,\cite{Doe97prb}, $0.3$-$0.7$\,km/s for $\beta$-W and NbReN\,\cite{Hof21tsf,Cir21prm}, $0.2$\,km/s for NbN\,\cite{Lin13prb} yield $\tau_\epsilon$ on the (sub-)ns time scale.

Recently, it was predicted theoretically and confirmed experimentally that FFI does not necessarily occur in the entire volume of the sample, but it can have a local character\,\cite{Bez19prb}. If the sample is inhomogeneous, then in the flux-flow regime it could be that regions with slower and faster moving vortices coexist, leading to the formation of normally conducting domains in the regions where $v^\ast$ is reached first. Once the current density exceeds some threshold current density (determining the equilibrium of the nonisothermal normal/superconducting boundary)\,\cite{Gur84spu,Bez84ltp} these domains begin to grow and the whole sample transits to the normal
state\,\cite{Bez19prb}. In this case, the standard relation $v^\ast = V^\ast/(B L)$ [$V^\ast$: measured voltage just before the transition, $B$: magnitude of the applied magnetic field, $L$: distance between the voltage leads] can no longer yield the instability velocity quantitatively. Nevertheless, the functional relations between the LO instability parameters in the case of a local FFI remain the same as in the case of a global FFI, but require renormalization of the sample length\,\cite{Bez19prb}. At the same time, it is worth noting that regardless of the FFI character, the places where FFI is actually nucleating are determined by defects at that edge through which vortices enter the sample\,\cite{Vod19sst}. Microscopically, such defects affect the vortex motion primarily via the local suppression of the edge barrier for vortex entry either due to variation of material parameters (critical temperature, electron mean free path etc.) or due to the local enhancement of the transport current density (current-crowding effect) near geometrical defects\,\cite{Buz98pcs,Ala01pcs,Vod03pcs,Cle11prb,Mik21prb}.

Here, we compare the non-equilibrium state generated by vortex motion in MoSi strips with focused ion beam (FIB)-milled and laser-etched edges. We choose MoSi because it is known for its very weak intrinsic pinning and as a material which is promising for single-photon detectors\,\cite{Cal18apl,Kor20pra,Cha20apl,Kor14sst}. Specifically, for MoSi strips with smooth edges milled by FIB we reveal that the critical current $I_\mathrm{c}$ is controlled by the edge barrier in a wide range of magnetic fields and $I_\mathrm{c}$ is close to the depairing current $I_\mathrm{dep}$ at $B=0$, which is indicative of a high homogeneity of the material. From the $I$-$V$ curves we deduce instability velocities $v^\ast$ in the $5$-$20$\,km/s range and energy relaxation times $\tau_\epsilon$ on the $20$\,ps time scale. By contrast, in the laser-etched MoSi strip, $I_\mathrm{c}$ is several times smaller than $I_\mathrm{dep}$, $I_\mathrm{c}(B)$ does not show indications for edge-barrier effects and the $v^\ast$ values are an order of magnitude smaller and close to the values deduced in the past for MoSi\,\cite{Doe97prb,Sam95prl} and other dirty superconductor films\,\cite{Hof21tsf,Liu21sst,Lin13prb}. We explain these differences by the change in the quality of the close-to-edge regions by the laser etching process which leads to a strong local inhomogeneity, bulk pinning and, thus, to a much lower instability velocity. Our results imply that for strips where no care was taken about the edge quality and bulk homogeneity, and whose magnetic field dependence of the critical current does not attest to the dominating edge-barrier effect, FFI only allows for the deduction of some ``indicative'' relaxation time exceeding the intrinsic $\tau_\epsilon$ in the material.

\section{Experiment}
\begin{figure}[t!]
    \centering
    \includegraphics[width=1\linewidth]{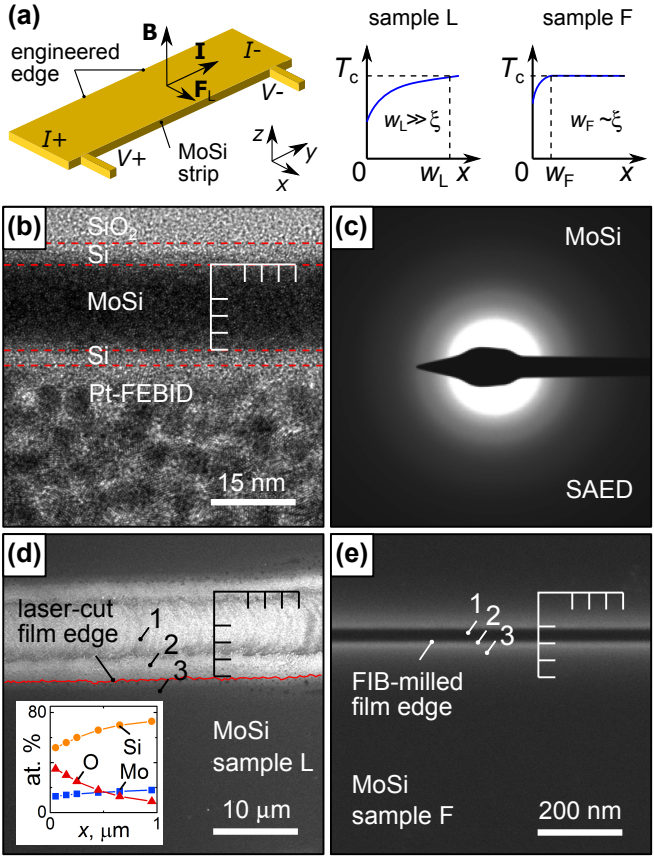}
    \caption{(a) Experimental geometry. The different quality of the edges creates different conditions for the entry of Abrikosov vortices into the strips.
    (b) TEM image and (c) selected area electron diffraction pattern of the MoSi film.
    (d) SEM images of the film regions near the edges formed by laser beam etching and (e) focused ion beam milling.
    The laser-etched edge is rough (red line in (d)) while the FIB-milled edge is smooth (e).
    Note the different scale in (d) and (e). Inset in (d): Material composition near the edge ($x=0$) in sample L.}
    \label{f1}
\end{figure}

The non-equilibrium state generated by fast vortex motion is investigated for two 15\,nm-thick amorphous superconducting MoSi strips differing by the edge roughness. The experimental geometry is shown in Fig.\,\ref{f1}(a). The MoSi films were deposited by dc magnetron co-sputtering of elemental molybdenum and silicon targets onto Si wafers covered with a thermally grown 230-nm-thick SiO$_2$ layer. The Mo$_{70}$Si$_{30}$ composition of the films was ensured by using the calibrated deposition rates and inspection of thicker film replica by energy-dispersive x-ray (EDX) spectroscopy. The films were deposited onto 5\,nm-thick Si buffer layers and covered with 3\,nm-thick Si layers for protection against oxidation. The films have a flat morphology, with an rms surface roughness of less than $0.1$\,nm, as deduced from atomic force microscopy (AFM) scans in the range $1\times1\,\mu$m$^2$. Microstructural characterization of MoSi lamellas by transmission electron microscopy (TEM: Tecnai F30, 300\,kV) revealed high structural uniformity of the films, see Fig. \ref{f1}(b). An exemplary selected area electron diffraction pattern of the MoSi film is shown in Fig.\,\ref{f1}(c). The absence of diffraction rings in Fig.\,\ref{f1}(c) attests to an amorphous microstructure of the material, in contrast with various polycrystalline microstructures\,\cite{Dob15bjn,Por19acs}.

For electrical resistance measurements the films were patterned into a four-probe geometry, with a strip length $L=616\,\mu$m and width $W=182\,\mu$m. One strip was etched out by laser beam (sample L) and another strip by Ga FIB milling (sample F). The use of different patterning techniques allowed for the realization of rough and smooth edges, respectively, see also Fig.\,\ref{f1}(d) and (e).

Laser etching was done under ambient conditions using an LGI-505 gas laser source, with $337$\,nm wavelength, $7$\,ns pulse duration and up to 1000 pulses per second of the laser. The beam power, focal spot size and speed of the beam rastering are decisive for the edge quality. These parameters were adjusted to produce an edge shown in Fig.\,\ref{f1}(d). The discrete spatial character of the laser beam impact is seen as a circle-footprint contrast variation along the groove etched in the substrate (region 1 in Fig.\,\ref{f1}(d)). Gaussian flanks of the laser beam, with a focal spot diameter of about $6\,\mu$m, caused evaporation of the MoSi film within a region of width
$2$-$3\,\mu$m along the edge (region 2). As a result, an irregular, saw-tooth-like strip edge profile (red thin line in Fig.\,\ref{f1}(d)) was created in sample L, characterized by an irregular variation of the edge barrier for vortex entry into the interior of the strip (region 3). The variation of the composition of sample L induced by the laser etching process within a region of width $w_\mathrm{L} \simeq 1\,\mu$m along the edge, is shown in the inset in Fig.\,\ref{f1}(d). The local film composition was inferred from EDX spectroscopy at $5$\,kV/$1.6$\,nA for a series of $100\times100$\,nm$^2$ areas probed at different distances from the strip edge. The larger oxygen content in the close-to-edge region leads to a degradation of the superconducting properties. The larger relative contents of Si and O with respect to the Mo$_{70}$Si$_{30}$ composition are because of the significantly larger thickness (of about $90$\,nm) of the layer probed by $5$\,kV electrons than the film thickness.

FIB milling was done in a dual-beam scanning electron microscope (SEM: FEI Nova NanoLab 600) at 30\,kV/30\,pA and 20\,nm pitch. The milling of a groove (region 1 in Fig.\,\ref{f1}(e)) was accompanied by stopping of Ga ions within a region of width $w_\mathrm{F}\sim10$\,nm along the edges in sample F, as inferred from SRIM simulations and seen as a lighter region 2 in the SEM image in Fig.\,\ref{f1}(e). The rms edge roughness in the $y$-direction is less than $0.5$\,nm, as deduced from an AFM scan over a distance of $500\,$nm along the edge. Thus, the edges of sample F produce a close-to-perfect edge barrier for vortex entry into the strip (region 3).

\begin{figure}[t!]
    \centering
    \includegraphics[width=1\linewidth]{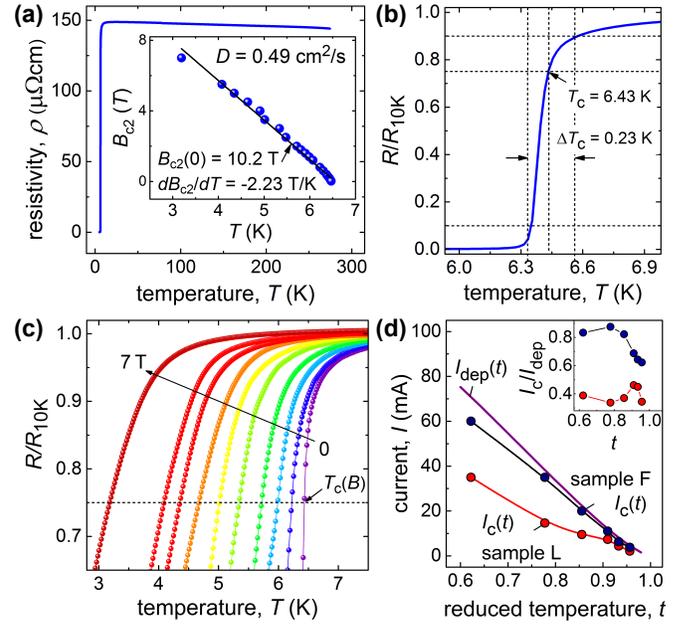}
    \caption{Superconducting properties of the films. (a) Temperature dependence of the resistivity of sample L. Inset: Temperature dependence of the upper critical field (symbols) os sample L fitted to the expression $B_\mathrm{c2}(T) = B_\mathrm{c2}(0)/0.7-(dB_\mathrm{c2}/dT)T$ (solid line). (b) Normalized resistance of sample L in the region of the superconducting transition. (c) Evolution of the superconducting transition of sample L with increase of the magnetic field. (d) Temperature dependence of the experimentally measured critical current $I_\mathrm{c}$(t) (symbols) for both samples in comparison with the Ginzburg-Landau depairing current calculated by Eq.\,\eqref{e1} (solid line). Inset: ratio $I_\mathrm{c}/I_\mathrm{dep}$ for both samples versus reduced temperature $t$.}
    \label{f2}
\end{figure}
Electrical resistance measurements were done in a He bath cryostat equipped with a superconducting solenoid. Magnetic field was applied perpendicular to the strip plane and the current-voltage ($I$-$V$) curves were recorded in the current-driven regime. The temperature dependence of the resistance exhibits a weak localization behavior (see Fig.\,\ref{f2}(a) which is representative for both samples), with a resistivity of $\rho_{\mathrm{7\,K}}\approx148\,\mu\Omega$cm at $7$\,K. Figure
\ref{f2}(b) depicts the superconducting transition of the MoSi film at $T_\mathrm{c} = 6.43$\,K, as determined by using the $75$\% resistance criterion, and the superconducting transition width $\Delta T_\mathrm{c} = 0.23$\,K, determined as the temperature interval between the $10$\% and $90$\% of the resistance at $7$\,K. Application of a magnetic field leads to a broadening of the transition, accompanied with its systematic shift toward lower temperatures, see Fig.\,\ref{f2}(c). Near $T_\mathrm{c}$, the temperature dependence of the upper critical field $B_\mathrm{c2}(T)$ exhibits a slope $dB_{\mathrm{c}2}/dT =-2.23$\,T/K, whose extrapolation toward zero temperature yields $B_\mathrm{c2}(0) \approx 10.2$\,T, see the inset in Fig.\,\ref{f2}(a). This slope corresponds to an electron diffusion coefficient $D$ of $0.49$\,cm$^2$/s, as deduced from the relation
$D = -1.097(dB_{\mathrm{c}2}/dT)^{-1}|_{T = T_\mathrm{c}}$\,\cite{Sem09prb}. The coherence length and the penetration depth at zero temperature are estimated\,\cite{Kor18pra} as $\xi(0) = \sqrt{\hbar D
/1.76k_\mathrm{B}T_\mathrm{c}} = 5.9$\,nm and $\lambda(0) = 1.05\cdot10^{-3} \sqrt{\rho_\mathrm{7K} /T_\mathrm{c}} \approx 495\,$nm, with the Pearl length $\Lambda = 2\lambda^2(0)/d \approx 32\,\mu$m. Thus, our strips are thin and wide, with $d \ll \lambda(0)$ and $\xi \ll \Lambda \lesssim W$.

\section{Results and Discussion}
\subsection{Critical current and current-voltage curves}
The maximal value of the dissipation-free current $I_{\mathrm{c}}$ the superconductor can carry is of primary importance for both, its use in single-photon detectors and the realization of ultra-fast vortex motion. Theoretically, at zero magnetic field, this current is given by the pair-breaking current $I_\mathrm{dep}$, whose temperature dependence can be described by the expression
\begin{equation}
\label{e1}
\begin{array}{lll}
    I_\mathrm{dep}(T) = I_\mathrm{dep}(0) (1 - (T/T_\mathrm{c})^{2})^{3/2},\quad{\mathrm{where}}\\
    \\
    I_\mathrm{dep}(0) = \displaystyle\frac{0.74 W [\Delta(0)]^{3/2}}{e R_\square \hbar D\sqrt{1+W/(\pi\Lambda)}}
\end{array}
\end{equation}
for dirty superconductors\,\cite{Rom82prb,Cle12prb,Kor18pra}. In Eq.\,\eqref{e1}, $\Delta(0)$ is the superconducting gap at zero temperature, $e$ the electron charge, and $R_\square$ the sheet resistance. The factor $\sqrt{1/(1+W/(\pi\Lambda))}$, which is absent for narrow strips ($W \ll \Lambda$), takes into account the nonuniform current distribution in strips with $W\gtrsim \Lambda$\,\cite{Plo01prb}. With the BCS ratio $\Delta(0)
\approx 1.76 k_\mathrm{B}T_\mathrm{c}$ we obtain $I_\mathrm{dep}(0)\approx 144\,$mA for our MoSi strips.

\begin{figure}[t!]
    \centering
    \includegraphics[width=1\linewidth]{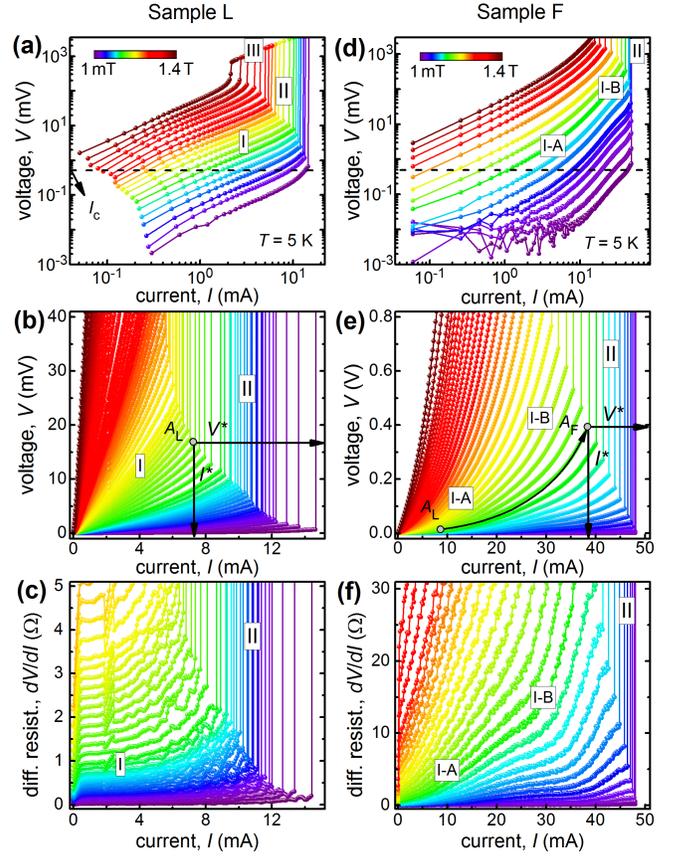}
    \caption{$I$-$V$ curves of the MoSi strips with laser-etched edges [sample L, panels (a)-(c)] and FIB-milled edges [sample F, panels (d)-(f)] at $T = 5$\,K for a series of magnetic fields. (b,e) Regimes of the nonlinear conductivity (I) and the flux-flow instability (II). The instability voltage $V^\ast$ and current $I^\ast$ are indicated for the $I$-$V$ curves at $B = 69$\,mT. (c,f) Current dependences of the differential resistances.}
    \label{f3}
\end{figure}

The theoretical dependence $I_\mathrm{dep}(T)$ calculated by Eq.\,\eqref{e1} is compared with the experimentally measured $I_\mathrm{c}(T)$ in Fig.\,\ref{f2}(d). We used the $0.5$\,mV voltage criterion for the deduction of the critical current $I_\mathrm{c}$ from the $I$-$V$ curves, as illustrated in Fig.\,\ref{f3}. This criterion corresponds to the lowest voltage at the foot of the zero-field resistance jump for sample F. Note that in the investigated temperature range $0.5<t<1$, where $t = T/T_\mathrm{c}$ is the reduced temperature, $I_\mathrm{c}$ varies between $0.3I_\mathrm{dep}$ and $0.5I_\mathrm{dep}$ for sample L and it is between $0.6I_\mathrm{dep}$ and $0.9I_\mathrm{dep}$ for sample F, see the inset in Fig.\,\ref{f2}(d).

The $I$-$V$ curves for both samples are presented in Fig.\,\ref{f3} for $T = 5$\,K ($0.78T_\mathrm{c}$). In panels (a) and (d) of Fig.\,\ref{f3} we label three distinct regimes in the $I$-$V$ curves: (I) the flux-flow regime, (II) the FFI, and (III) the normal conducting regime. Panels (b) and (e) of Fig.\,\ref{f3} show in more detail the nonlinear conductivity regimes preceding the voltage jumps. The last points (A) before the jumps correspond to the instability current $I^\ast$ related to the instability voltage $V^\ast$.

A comparison of the $I$-$V$ curves for both samples suggests that sample L transits into the highly-resistive state at noticeably smaller currents than sample F. Herewith, nonlinear upturns in the $I$-$V$ curves at the foot of the instability jump occur in a broader range of currents for sample F. This behavior is illustrated by the evolution of the instability point $A_\mathrm{L}$ for sample L to the instability point $A_\mathrm{F}$ for sample F for the $I$-$V$ curves taken at the same field of $69$\,mT in panels (b) and (e). The extended regime of nonlinear conductivity (I-B) for sample F is also seen in the $dV/dI$ versus $I$ representation in Fig.\,\ref{f3}(c) and (f), as compared to the regime of almost linear conductivity (regime I for sample L and regime I-A for sample F) at smaller currents. In this way, at a given magnetic field magnitude, sample F exhibits larger $V^\ast$ than sample L. This enhancement is most pronounced at low magnetic fields.

\subsection{Field dependence of the critical current}
Magnetic-field dependence of the critical current allows for the identification of various states the superconductor is passing with increase of the magnetic field\,\cite{Plo01prb,Ili14prb,Dob20nac}. Namely, at low magnetic fields the sample can be in the vortex-free (Meissner) state, resulting in a linear decrease of $I_\mathrm{c}(B)$. At higher fields, the decrease slows down, with
a crossover at $B_{\mathrm{stop}}$ demarcating a transition to the mixed state\,\cite{Mak98pss}. In the Meissner state ($B < B_\mathrm{stop}$),
\begin{equation}
\label{e2}
    \begin{array}{lll}
    I_\mathrm{c}(B) = I_\mathrm{c}(0\,\mathrm{T})(1- B/2B_\mathrm{stop}),\quad \mathrm{where}\\
    \\
    B_\mathrm{stop} = \displaystyle\frac{\Phi_0 \sqrt{1+W/(\pi\Lambda)}}{2\sqrt3\pi\xi(T) W}
    \end{array}
\end{equation}
contains the factor $\sqrt{1+W/(\pi\Lambda)}$ because of the width $W \gtrsim \Lambda$. Equation\,\eqref{e2} allows us to estimate $B_\mathrm{s} = 2B_\mathrm{stop}$ as $0.3$\,mT for sample F. The physical meaning of $B_\mathrm{s}$ is the field value at which the surface barrier for vortex entry is suppressed at $I = 0$. The magnetic field dependence of the critical current for both samples is presented in Fig.\,\ref{f4}(a). One can see that for sample F the theoretically calculated $B_\mathrm{stop} \sim 0.15$\,mT is not very far from the experimental $B_\mathrm{stop} \sim 0.2$\,mT and that the very steep decrease of $I_\mathrm{c}(B)$ for sample F at $B\rightarrow0$ (see the inset in Fig.\,\ref{f4}(a)) could be an indication for a vortex-free state. Furthermore, the value of $I_\mathrm{c}(0)$ for sample F is a factor of 3 larger than $I_\mathrm{c}(0)$ for sample L, attesting to strong edge-barrier effects. For 1\,mT$\leq B\leq100$\,mT, $I_\mathrm{c}(B)$ for sample F is described well by the dependence $I_\mathrm{c}(B) =
I_\mathrm{c}(0\,\mathrm{T})B_\mathrm{stop}/2B$, indicating the dominating role of the edge barrier for vortex entry in sample F at $B\lesssim100$\,mT. In general, one could expect a slowing down towards the dependence $I_\mathrm{c}(B)\sim B^{-0.5}$ in larger fields\,\cite{Dob20nac} because of the transition to the regime of dominating intrinsic (volume) pinning. Indeed, such a transition occurs at larger fields ($\backsimeq300$\,mT, not shown) in sample F, pointing to the very small contribution of bulk pinning. This finding is in line with the high structural homogeneity of the amorphous MoSi films (see also Fig.\,\ref{f1}(b) and (c)).

By contrast, in sample L, the dependence $I_\mathrm{c}(B)$ is flattened at $B\lesssim 5$\,mT. This behavior cannot be explained by a reduction of the edge barrier since the edge roughness leads only to a decrease of $I_\mathrm{c}(0)$ and $B_\mathrm{stop}$ but not to the change of the functional dependence $I_\mathrm{c}(B)$. We believe that this behavior is connected with the change of the structure of MoSi by the laser-beam impact, within a distance of about $2\,\mu$m from the edge, leading to a strong inhomogeneity of MoSi near the edge and the appearance of vortex pinning there. Some indication for the vortex pinning comes not only from the drastically different dependence of $I_\mathrm{c}(B)$ in sample L in comparison with sample F, but also from the comparison of their $I$-$V$ curves. Namely, Fig.\,\ref{f3}(a) reveals an exponential shape of the $I$-$V$ curves which could be considered as an indication of the vortex creep in the near-edge region of sample L.

\subsection{Maximal vortex velocity}
\begin{figure}[t!]
    \centering
    \includegraphics[width=1\linewidth]{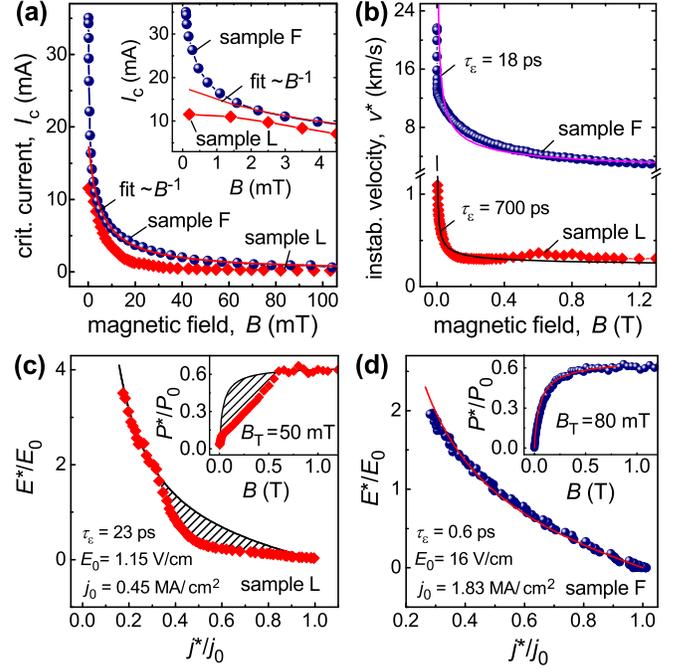}
    \caption{(a) $I^\ast(B)$ for both samples. Symbols: experiment; line: fit to Eq.\,\eqref{e2}. Inset: same $I_\mathrm{c}(B)$ data at low fields.
    (b) $v^\ast(B)$ for both samples. Symbols: experiment; lines: fits to Eq.\,\eqref{eHub} with the energy relaxation time used as the only fitting parameter. (c) and (d) Fits of the experimental data (symbols) to Eq.\,\eqref{eLOparam} (lines). The hatched area in (c) emphasizes the range of deviation of the experimental data for sample L from the calculation. In all panels $T = 5$\,K.}
    \label{f4}
\end{figure}

The magnetic field dependence of the maximal vortex velocity $v^\ast$, deduced by using the standard relation $v^\ast=V^\ast/(BL)$, is presented in Fig. \ref{f4}(b). For both samples $v^\ast(B)$ decreases with increase of $B$,
roughly following the $v^\ast\sim B^{-1/2}$ law. However, the $v^\ast$ values differ substantially, with $v^\ast$ for sample F being a factor of $\thicksim20$ larger than for sample L in the whole range of magnetic fields. For instance, for sample F, $v^\ast$ reaches $\thicksim13\,$km/s at $5$\,mT, decreases to $\thicksim7$\,km/s at $200$\,mT, and then slowly decreases to $\thicksim3$\,km/s at $1$\,T. At the same time, for sample L, $v^\ast\approx800$\,m/s at $5$\,mT, decreases to $300$\,m/s at $200$\,mT and remains nearly constant with a further increase of the magnetic field.

According to LO, the instability velocity $v^\ast$ is independent of the magnetic field. As was argued by Doettinger\,\emph{et al}\,\cite{Doe95pcs}, the discrepancy between the field-dependent experimental $v^\ast$ and the field-independent theoretical $v^\ast$ can be overcome by taking into account the magnetic field dependence of the vortex lattice parameter. Namely, the non-equilibrium electron distribution is spatially uniform only whilst $v^\ast\tau_\mathrm{\epsilon}$ is larger than the intervortex distance $a$. This regime is realized at high magnetic fields, as is also in line with our data in Fig.\,\ref{f4}(b), where an almost constant $v^\ast$ is observed at $B \gtrsim 600$\,mT. At smaller fields the system can be recovered to a spatially homogeneous state by allowing $v^\ast$ to grow accordingly to the increase of $a$ with decrease of the applied magnetic field, $a=\sqrt{2\Phi_0/\sqrt{3}B}$, where $\Phi_0$ is the magnetic flux quantum. The LO expression is then complemented\,\cite{Doe95pcs} with the term $a/\sqrt{D \tau_\epsilon}$, yielding
\begin{equation}
\label{eHub}
    v^\ast = \left[\frac{(1-t)^{1/2}D[14\zeta(3)]^{1/2}}{\pi \tau_\epsilon}\right]^{1/2}
    \left(1 + \frac{a}{\sqrt{D \tau_\epsilon}}\right).
\end{equation}

The fits of our experimental data to Eq.\,\eqref{eHub} are shown by solid lines in Fig.\,\ref{f4}(b). Herewith, $v^\ast(B)$ is calculated while varying the energy relaxation time $\tau_\epsilon$ as the only fitting parameter. Specifically, the fits shown in Fig.\,\ref{f4}(b) were obtained with $\tau_\epsilon = 700$\,ps and $19$\,ps for sample L and sample F, respectively. The ratio $\tau_{\varepsilon\,L}/\tau_{\varepsilon\,F}\approx40$ illustrates that the fits by Eq.\,\eqref{eHub}, which is widely used in recent works \cite{Liu21sst,Hof21tsf}, yield $\tau_\epsilon$ differing by more than an order of magnitude for strips made from the same material, but differing by the edge quality.

\subsection{Larkin-Ovchinnikov-Bezuglyj-Shklovskij (LOBS) model}
The LO theory was generalized by Bezuglyj and Shklovskij (BS)\,\cite{Bez92pcs} for a finite rate of heat removal from the superconductor to the substrate. Based on the heat balance equation, BS introduced a new field parameter, the overheating field $B_\mathrm{T}$
\begin{equation}
\label{eBt}
    B_\mathrm{T} = 0.374 k_\mathrm{B}^{-1}eR_{\square}h\tau_\epsilon,
\end{equation}
where $k_\mathrm{B}$ is the Boltzmann constant and $h$ the heat removal coefficient. The parameter $B_\mathrm{T}$ separates the region of small fields $B\lesssim B_\mathrm{T}$ at which heat removal is fast enough and the instability is of non-thermal nature from the region of large fields $B_\mathrm{T} \lesssim B \lesssim 0.4 B_\mathrm{c2}$ with insufficient heat removal and the heating mechanism dominating the instability. BS derived a scaling law for the electric field strength $E^\ast$ and the current density $j^\ast$ at the instability point
\begin{equation}
\label{eLO}
    \displaystyle\frac{E^\ast}{E_0}= (1 - f(b))\left(\frac{j^\ast}{j_0}\right)^{-1},
\end{equation}
where $E_0$, $j_0$ and $f$ are defined as
\begin{equation}
\label{eLOparam}
    \begin{array}{lll}
    \displaystyle E_0 = 1.02 B_\mathrm{T}(D/\tau_\epsilon)^{1/2}(1-t)^{1/4},
    \\[2mm]
    \displaystyle j_0 = 2.62 (\sigma_\mathrm{n}/e)(D\tau_\epsilon)^{-1/2}k_\mathrm{B} T_\mathrm{c}(1-t)^{3/4},
\\[2mm]
    \displaystyle f =[1+b+(b^2 +8b+4)^{1/2}]/3(1+2b),
    \end{array}
\end{equation}
with the reduced magnetic field $b = B / B_\mathrm{T}$ and the normal state conductivity $\sigma_\mathrm{n}$. If one fits the entirety of the experimentally deduced instability points $(E^\ast/E_0)(j^\ast/j_0)$ to Eq.\,\eqref{eLOparam}, then the field dependence of the power density at the instability point, $P_0\equiv E_0 j_0 = h/d (T_\mathrm{c}-T)$ allows one to deduce the overheating field $B_\mathrm{T}$ and the heat removal coefficient $h$. Substitution of $h$ and $B_\mathrm{T}$ into Eq.\,\eqref{eBt} yields then the relaxation time $\tau_\epsilon$.

The best fits of the experimental data to Eq.\,\eqref{eLOparam} for both samples are presented in Fig.\,\ref{f4}(c) and (d). For sample L the fit is very poor since in the range of currents $0.4\lesssim (j^\ast/j_0)\lesssim 0.8$ the experimental data strongly deviate below the curve calculated by Eq.\,\eqref{eLOparam}, see the hatched area in Fig.\,\ref{f4}(c). The fit in Fig.\,\ref{f4}(c) was done with $E_0 = 1.15$\,V/cm and $j_0 =0.45$\,MA/cm$^2$, yielding $B_\mathrm{T} = 50$\,mT, $h = 0.54$\,W/Kcm$^2$, and $\tau_\varepsilon=23$\,ps. For sample F the fit is almost perfect as the experimental data fall onto the curve calculated by Eq.\,\eqref{eLOparam} in the entire range of magnetic fields. The fit in Fig.\,\ref{f4}(d) was done with $E_0 = 16$\,V/cm and $j_0 =1.83$\,MA/cm$^2$, yielding $B_\mathrm{T} = 80$\,mT, $h = 32.7$\,W/Kcm$^2$, and $\tau_\varepsilon=0.6$\,ps. If one associates $\tau_{\epsilon}$ with the electron-phonon scattering time $\tau_\mathrm{ep}$ in the LO model\,\cite{Lar86inb}, the deduced $\tau_{\epsilon}$ is at least one order of magnitude smaller than one could expect from $\tau_{\epsilon}$ found in similar low-$T_\mathrm{c}$ highly disordered superconductors\,\cite{Bab04prb,Sid18prb,Sid20prb}. We note that the LO theory was developed neglecting the diffusion term in the kinetic equation. This approximation is justified when the intervortex distance $a$ is smaller than the quasiparticles diffusion length $l_\varepsilon = \sqrt{D\tau_\varepsilon}$, with $l_{\varepsilon\,L} \approx 34$\,nm and $l_{\varepsilon\,F} \approx 7$\,nm suggesting that the deduced $\tau_\epsilon$ values can hardly be treated even as order-of-magnitude estimates. Nonetheless, while the values of $\tau_\varepsilon$ deduced from the LOBS model\,\cite{Bez92pcs} differ from those deduced from the model of Doettinger\,\emph{et al}\,\cite{Doe95pcs}, the \emph{ratio} between the deduced relaxation times is almost the same, $\tau_{\varepsilon\,\mathrm{L}}/\tau_{\varepsilon\,\mathrm{F}}\approx40$.

\subsection{Numerical modeling}
To get further insights into the spatiotemporal evolution of the order parameter in the strips with and without edge defects, we numerically solve the modified TDGL equation in conjunction with the heat balance equation. The essential equations and the considered boundary conditions are detailed in Appendix A.

The shapes of the edges of sample L are unknown and their exact modeling is not feasible with the currently available computation capabilities. Therefore we consider the effects of single edge defects of different shapes and an array of defects located near the edge of the strip. The defects are simulated as regions with a locally suppressed critical temperature.
\begin{figure}[t!]
    \centering
    \includegraphics[width=1\linewidth]{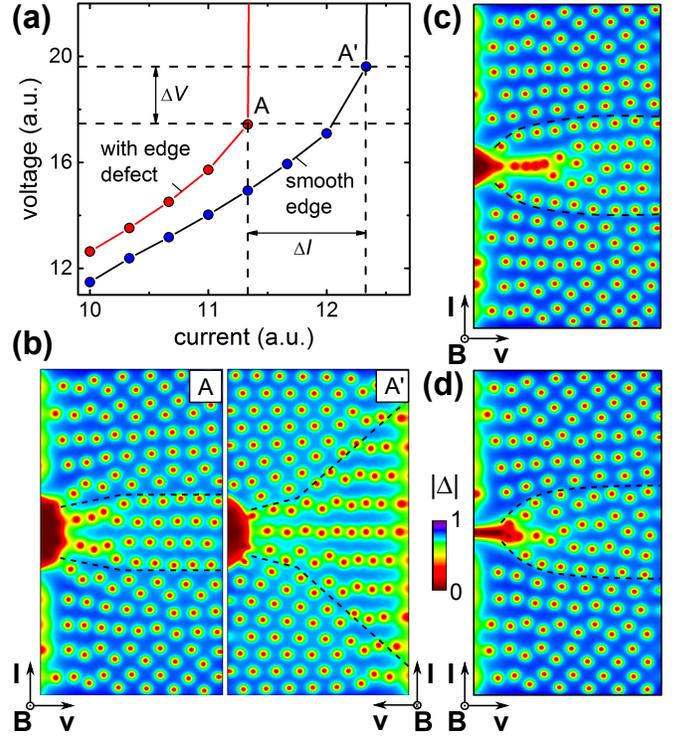}
    \caption{(a) $I$-$V$ curves calculated by the numerical solution of the TDGL for a superconducting strip with width $W = 100\xi_\mathrm{c}$ at $T = 0.8T_\mathrm{c}$ for $B = 0.05\,B_\mathrm{0}$ for the vortex entry via the edge with a defect and via the (opposite) smooth edge. (b) Snapshots of the superconducting order parameter $|\Delta|$ are shown at the different current values $I/I_\mathrm{dep} = 0.34$ (point A) and $I/I_\mathrm{dep} = 0.37$ (point A$^\prime$), corresponding to the instability points in panel (a). Dashed lines encage the regions of nucleation and evolution of vortex rivers. (c,d) Nucleation of vortex jets (branched vortex rivers) upon vortex entry through an edge defect. For the studied system, the parameter $B_0 = \Phi_0/(2\pi\xi_\mathrm{c}^2) \simeq 4.9$\,T, where $\xi_\mathrm{c}= \sqrt{1.76}\xi(0) = 7.8$\,nm.}
    \label{f5}
\end{figure}

The results of TDGL simulations for a single defect are presented in Fig.\,\ref{f5}. Figure\,\ref{f5}(a) presents the $I$-$V$ curves calculated for the vortex entry through the edge containing a single, semicircle-shaped edge defect in comparison with the $I$-$V$ curve for the vortex entry through the perfectly straight edge of the same strip. The simulations suggest that the low-dissipative regime of nonlinear conductivity extends toward larger currents (and, hence, higher vortex velocities) for the vortex entry via the perfect edge (point A$^\prime$ Fig.\,\ref{f5}(a)). The spatial dependences of the superconducting order parameter at the last points before the
instability jumps are illustrated in Fig.\,\ref{f5}(b). Panel A in Fig.\,\ref{f5}(b) illustrates the case when vortices enter via the edge containing a semicircle-shaped defect, while panel A$^\prime$ illustrates the case when they enter via the opposite smooth edge. The influence of the edge defect on the vortex dynamics is twofold. First, due to the current-crowding effect the defect suppresses the edge barrier for the penetration of vortices, turning the defect into the place of nucleation of vortex rivers\,\cite{Vod19sst}. Such vortex rivers represent self-organized Josephson-like junctions formed by chains of fast-moving vortices, which eventually evolve to normal domains expanding across the entire sample upon its abrupt transition to the highly-resistive state. Second, the local deflection of the current flow direction from the $y$ axis (see Fig.\,\ref{f1}(a)) near the defect leads to the deviation of the Lorentz force direction and, hence, the direction of vortex motion from the $x$ axis. As a result, a diverging ``jet'' of vortices is formed, as indicated by the dashed lines in panel A in Fig.\,\ref{f5}(b).

By contrast, when vortices enter via the perfect edge (panel A$^\prime$ in Fig.\,\ref{f5}(b)), the only effect which remains in comparison with the previously considered case is the local enhancement of the current density near the edge defect. Panel A$^\prime$ in Fig.\,\ref{f5}(b) shows that the nucleation of vortex rivers occurs at many different points along the perfect edge, while only those nucleated at the edge in front of the defect develop faster because of the larger current density. Figure\,\ref{f5}(c) and (d) illustrate the development of vortex ``jets'' upon vortex entry via the edge containing a triangle-shaped and a slit-shaped defect.

The simulations also allow us to explain the role of a single defect on the FFI observed in Ref.\,\cite{Dob20nac}. As in that experiment, our model gives a several-percent suppression of $I^\ast$ and $V^\ast$ when vortices enter via an edge of the strip containing a defect compared to entry through the smooth edge, while $I_\mathrm{c}$ can be suppressed by more than two times (depending on the size and shape of the defect). That experimental observation\,\cite{Dob20nac} follows from the considered model, where FFI starts near the edge because of the higher local current density $j$ and the electronic temperature $T_\mathrm{e}$, but in the rest of the strip the vortices have to move at a large enough velocity to allow for the appearance of vortex rivers\,\cite{Sil10prl}. Thus, the edge defect increases $j$ and $T_\mathrm{e}$ in comparison with the strip with the smooth edge and it leads to smaller $I^\ast$ and $V^\ast$, but their values do not change drastically  since the change in the current distribution decays fast with increase of the distance from the defect (approximately inversely proportional to the square of the distance).

In our experiment, the difference in $I^\ast$ and $V^\ast$ for samples L and F is several times larger. In comparison with Ref.\,\cite{Dob20nac} in our case the laser etching not only creates edge defects but it also changes the material composition of the close-to-edge regions. To take this into account in the simulations, we introduced randomly distributed defects (each defect has a size of $2\xi_\mathrm{c}\times2\xi_\mathrm{c}$ with a locally suppressed $T_\mathrm{c}$) in the close-to-edge region of width $25 \xi_\mathrm{c}$ for the strip with $W=100 \xi_\mathrm{c}$. However, the results for this system are similar to the ones for the strip with a single edge defect, namely, a relatively small suppression of $I^\ast$ and $V^\ast$ while the critical current could be suppressed significantly (as in the case of a single defect).

A comparison of the experimental results for samples L and F implies that FFI appears first in the close-to-edge region of sample L with width $2\,\mu$m\,$\ll182\mu$m (which is inhomogeneous due to the laser etching) and only then the FFI spreads to the rest of the superconducting strip. Unfortunately, our model does not describe this process. We also tried to simulate the initial FFI by introducing a local increase of the escape time $\tau_\mathrm{esc}$ of nonequilibrium phonons to the substrate in the close-to-edge region. Specifically, $\tau_\mathrm{esc}$ was larger by an order of magnitude in the region with width $25 \xi_\mathrm{c}$ for a strip with $W=100 \xi_\mathrm{c}$. Indeed, we have found that in that region FFI sets on first but it does not spread deep into the strip, because of the cooling of the close-to-edge regions via diffusion of hot electrons to the neighboring regions. We have to note that the so-called healing length at the chosen parameters in our simulations is much larger than the length of penetration of the electric field. Whether it plays an important role or does not for spreading of the normal region into the interior of the strip should be clarified in further investigations.

\section{Conclusion}
To sum up, we have investigated the effects of edge quality on the critical current and maximal vortex velocities in wide thin films of MoSi. The edges of different quality were produced by laser etching (sample L) and milling by a focused ion beam (sample F). The smooth edges in sample F have allowed for (i) a factor of about 3 increase of the zero-field critical current, (ii) a factor of 20 enhancement of the maximal vortex velocity up to about 20\,km/s, and (iii) a factor of 40 smaller estimate for the energy relaxation time on the 20\,ps time scale.

Our results have following implications for superconducting devices. First, the enhancement of the current-carrying capability of the strips is relevant for superconductor microstrip single-photon detectors (SMSPDs). Namely, to achieve the theoretically predicted intrinsic detection efficiency of about 100\%, SMSPDs should be biased by close-to-deparing critical currents\,\cite{Vod17pra}. In particular, in the MoSi strips with FIB-milled edges at 5\,K the zero field critical current has increased by a factor of 3, reaching 87\% of the Ginzburg-Landau pair-breaking current. Second, the improvement of the current-carrying capability has allowed for the enhancement of the maximal vortex velocity, providing access to the previously inaccessible regimes (in this material) of generation of sound and spin waves via a Cherenkov-type mechanism\,\cite{Bul05prb,Bes14prb,Dob21arx}, with a rich physics of fluxon-phonon and fluxon-magnon interactions\,\cite{Ivl99prb,Dob19nph}. From the viewpoint of basic research, our findings could be relevant for a possible explanation of the inconsistency between the (sub-)ns relaxation times deduced from current-voltage measurements for many dirty superconductors, despite their potential (or already proven) capability of single-photon counting\,\cite{Liu21sst,Sam95prl,Doe97prb,Hof21tsf,Cir21prm,Lin13prb}.

In all, our findings suggest the edge quality check as a route to improvement of the critical current in superconducting microstrip single-photon detectors and imply that homogeneous, dirty-limit superconductors with weak pinning should generally allow for ultra-fast vortex motion at velocities exceeding $10$\,km/s. In particular, our results imply that for strips where no care was taken about the edge quality and the magnetic field dependence of the critical current does not attest to the dominating edge pinning of vortices, the flux-flow instability only allows for the deduction of some ``indicative'' relaxation time exceeding the intrinsic $\tau_\varepsilon$ in the material.

\section*{Appendix A}
Simulation results presented in Fig.\,\ref{f5} rely upon the solution of the modified TDGL equation\,\cite{Vod17pra}
\begin{eqnarray*}
    \frac{\pi\hbar}{8k_\mathrm{B}T_\mathrm{c}} \left(\frac{\partial }{\partial t}+\frac{2ie\varphi}{\hbar} \right) \Delta= \nonumber
    \\
    =\xi^2_\mathrm{mod}\left( \nabla -i\frac{2e}{\hbar c}A\right)^2\Delta+\left(1-\frac{T_\mathrm{e}}{T_\mathrm{c}}-\frac{|\Delta|^2}{\Delta_{mod}^2}\right)\Delta+
    \nonumber
    \\
    +i\frac{(\mathrm{div}\mathbf{j}_\mathrm{s}^\mathrm{Us}-\mathrm{div}\mathbf{j}_\mathrm{s}^\mathrm{GL})}{|\Delta|^2}\frac{e\Delta\hbar D}{\sigma_\mathrm{n}\sqrt{2}\sqrt{1+T_\mathrm{e}/T_\mathrm{c}}},
\end{eqnarray*}
where $\xi^2_\mathrm{mod}=\pi\sqrt{2}\hbar D/(8k_\mathrm{B}T_\mathrm{c}\sqrt{1+T_\mathrm{e}/T_\mathrm{c}})$, $A$ is the vector potential, $\varphi$ the electrostatic potential, $D$ the diffusion coefficient,
$\Delta_\mathrm{mod}^2=(\Delta_0\tanh(1.74\sqrt{T_\mathrm{c}/T_\mathrm{e}-1}))^2/(1-T_\mathrm{e}/T_\mathrm{c})$, $\sigma_\mathrm{n}=2e^2DN(0)$ the normal-state conductivity,
$N(0)$ the single-spin density of states at the Fermi level, and
$\mathbf{j}_\mathrm{s}^\mathrm{Us}$ and
$\mathbf{j}_s^\mathrm{GL}$ are the superconducting current densities in the Usadel and Ginzburg-Landau models
\begin{equation*}
    \mathbf{j}_\mathrm{s}^\mathrm{Us} = \frac{\pi\sigma_\mathrm{n}}{2e\hbar}|\Delta|
    \tanh\left(\frac{|\Delta|}{2k_\mathrm{B}T_\mathrm{e}}\right)\mathbf{q}_\mathrm{s},
\end{equation*}
where $\mathbf{q}_\mathrm{s} = \nabla \phi - 2e\mathbf{A}/\hbar c$, $\phi$ is the phase of $\Delta=|\Delta|e^{i\phi}$, and
\begin{equation*}
    \mathbf{j}_\mathrm{s}^\mathrm{GL} = \frac{\pi\sigma_\mathrm{n}|\Delta|^2}{4e\hbar k_\mathrm{B}T_\mathrm{c}}\mathbf{q}_\mathrm{s}.
\end{equation*}
At $T_\mathrm{e}$ not very close to $T_\mathrm{c}$ the Ginzburg-Landau expression for the superconducting current is not valid quantitatively and one needs to use the Usadel expression for $\mathbf{j}_\mathrm{s}^\mathrm{Us}$. In this case the TDGL equation should also be modified since the ordinary TDGL equation leads to $\mathrm{div}\mathbf{j}_\mathrm{s}^\mathrm{GL} = 0$ in the stationary case, while one needs $\mathrm{div}\mathbf{j}_\mathrm{s}^\mathrm{Us} = 0$. Accordingly, by adding the term $\mathrm{div}(\mathbf{j}_\mathrm{s}^\mathrm{Us} - \mathbf{j}_\mathrm{s}^\mathrm{GL})$ in the TDGL equation one provides $\mathrm{div}\mathbf{j}_\mathrm{s}^\mathrm{Us} = 0$. At $T_\mathrm{e}\rightarrow T_\mathrm{c}$ the modified TDGL equation reduces to the ordinary TDGL equation and $\mathrm{div}(\mathbf{j}_\mathrm{s}^\mathrm{Us} - \mathbf{j}_\mathrm{s}^\mathrm{GL})$ goes to zero.

The electron and phonon temperatures, $T_\mathrm{e}$ and $T_\mathrm{p}$, respectively, are found from the solution of following equations
\begin{eqnarray*}
    \frac{\partial}{\partial t}\left(\frac{\pi^2k_\mathrm{B}^2N(0)T_\mathrm{e}^2}{3}-
    \mathcal{E}_0\mathcal{E}_\mathrm{s}(T_\mathrm{e},|\Delta|)\right) = \nonumber
        \\
    = \nabla k_\mathrm{s} \nabla T_\mathrm{e}-\frac{96\zeta(5)N(0)k_\mathrm{B}^2}{\tau_0}\frac{T_\mathrm{e}^5-
    T_\mathrm{p}^5}{T_\mathrm{c}^3}+ j E,
        \\
    \frac{\partial T_\mathrm{p}^4}{\partial t}=-\frac{T_\mathrm{p}^4- T^4}{\tau_\mathrm{esc}}+\gamma\frac{24\zeta(5)}{\tau_0}\frac{15}{\pi^4}\frac{T_\mathrm{e}^5-T_\mathrm{p}^5}{T_\mathrm{c}},
\end{eqnarray*}
where $\mathcal{E}_0=4N(0)(k_\mathrm{B}T_\mathrm{c})^2$, $\mathcal{E}_0\mathcal{E}_\mathrm{s}(T_\mathrm{e},|\Delta|)$ is the change in the
energy of electrons due to the transition to the superconducting state, $k_\mathrm{s}$ is the heat conductivity in the superconducting state
\begin{equation*}
    k_\mathrm{s}=k_\mathrm{n}\left(1-\frac{6}{\pi^2(k_\mathrm{B}T_\mathrm{e})^3}\int_0^{|\Delta|}\frac{\epsilon^2
    e^{\epsilon/k_\mathrm{B}T_\mathrm{e}}d\epsilon}{(e^{\epsilon/k_\mathrm{B}T_\mathrm{e}}+1)^2}\right),
\end{equation*}
$k_\mathrm{n} = 2D\pi^2k_\mathrm{B}^2N(0)T_\mathrm{e}/3$ is the heat conductivity in the normal state, the term $jE$ describes Joule dissipation, and $\tau_\mathrm{esc}$ is the escape time of nonequilibrium phonons to the substrate. The parameter $\gamma$ is defined as
$\gamma= \displaystyle\frac{8\pi^2}{5}\displaystyle\frac{C_\mathrm{e}(T_\mathrm{c})}{C_\mathrm{p}(T_\mathrm{c})}$, where $C_\mathrm{e}(T_\mathrm{c})$ and $C_\mathrm{p}(T_\mathrm{c})$ are the heat capacities of electrons and phonons at $T=T_\mathrm{c}$, and the characteristic time $\tau_0$ controls the strength of the electron-phonon and phonon-electron scattering \cite{Vod17pra}. Values of the parameters $\gamma=9$ and $\tau_0=925$\,ns used in the calculations are estimates for NbN. Their variation only leads to quantitative changes in the $I$-$V$ curves.

The current continuity equation $\mathrm{div}(\mathbf{j}_\mathrm{s}^{Us}+\mathbf{j}_\mathrm{n})=0$ is solved to find the electrostatic potential. Here, $\mathbf{j}_\mathrm{n}=-\sigma_\mathrm{n}\nabla \varphi$ is the normal current density. At the edges where vortices enter and exit the microstrip we use the boundary conditions $\mathbf{j}_\mathrm{n}|_\mathrm{n}=\mathbf{j}_\mathrm{s}|_\mathrm{n}=0$ and $\partial T_\mathrm{e}/\partial\mathrm{n}=0$, $\partial |\Delta|/\partial n=0$ while at the edges along the current direction $T_\mathrm{e}=T$, $|\Delta|=0$, $\mathbf{j}_\mathrm{s}|_\mathrm{n}=0$, $\mathbf{j}_\mathrm{n}|_\mathrm{n}=I/wd$. The latter boundary conditions model the contact of the superconducting strip with a normal reservoir being in equilibrium. This choice provides a way ``to inject'' the current into the superconducting microstrip in the modeling. The modeled length of the microstrip is $L=4w$.

\begin{acknowledgments}
The authors thank Roland Sachser for support with the nanofabrication. B.B. acknowledges financial support by the Vienna Doctoral School in Physics (VDSP). D.Y.V. acknowledges support by the Russian Foundation for Basic Research (RFBR), grant No. 18-29-20100. Support through the Frankfurt Center of Electron Microscopy (FCEM), by the Austrian Science Fund (FWF) grant No. I4865N and by the European Cooperation in Science and Technology via COST Actions CA16218 (NANOCOHYBRI) and CA19108 (HiSCALE) is gratefully acknowledged.
\end{acknowledgments}

\newpage


\begin{thebibliography}{61}%
\makeatletter
\providecommand \@ifxundefined [1]{%
 \@ifx{#1\undefined}
}%
\providecommand \@ifnum [1]{%
 \ifnum #1\expandafter \@firstoftwo
 \else \expandafter \@secondoftwo
 \fi
}%
\providecommand \@ifx [1]{%
 \ifx #1\expandafter \@firstoftwo
 \else \expandafter \@secondoftwo
 \fi
}%
\providecommand \natexlab [1]{#1}%
\providecommand \enquote  [1]{``#1''}%
\providecommand \bibnamefont  [1]{#1}%
\providecommand \bibfnamefont [1]{#1}%
\providecommand \citenamefont [1]{#1}%
\providecommand \href@noop [0]{\@secondoftwo}%
\providecommand \href [0]{\begingroup \@sanitize@url \@href}%
\providecommand \@href[1]{\@@startlink{#1}\@@href}%
\providecommand \@@href[1]{\endgroup#1\@@endlink}%
\providecommand \@sanitize@url [0]{\catcode `\\12\catcode `\$12\catcode
  `\&12\catcode `\#12\catcode `\^12\catcode `\_12\catcode `\%12\relax}%
\providecommand \@@startlink[1]{}%
\providecommand \@@endlink[0]{}%
\providecommand \url  [0]{\begingroup\@sanitize@url \@url }%
\providecommand \@url [1]{\endgroup\@href {#1}{\urlprefix }}%
\providecommand \urlprefix  [0]{URL }%
\providecommand \Eprint [0]{\href }%
\providecommand \doibase [0]{http://dx.doi.org/}%
\providecommand \selectlanguage [0]{\@gobble}%
\providecommand \bibinfo  [0]{\@secondoftwo}%
\providecommand \bibfield  [0]{\@secondoftwo}%
\providecommand \translation [1]{[#1]}%
\providecommand \BibitemOpen [0]{}%
\providecommand \bibitemStop [0]{}%
\providecommand \bibitemNoStop [0]{.\EOS\space}%
\providecommand \EOS [0]{\spacefactor3000\relax}%
\providecommand \BibitemShut  [1]{\csname bibitem#1\endcsname}%
\let\auto@bib@innerbib\@empty
\bibitem [{\citenamefont {Embon}\ \emph {et~al.}(2017)\citenamefont {Embon},
  \citenamefont {Anahory}, \citenamefont {Jelic}, \citenamefont {Lachman},
  \citenamefont {Myasoedov}, \citenamefont {Huber}, \citenamefont {Mikitik},
  \citenamefont {Silhanek}, \citenamefont {Milosevic}, \citenamefont
  {Gurevich},\ and\ \citenamefont {Zeldov}}]{Emb17nac}%
  \BibitemOpen
  \bibfield  {author} {\bibinfo {author} {\bibfnamefont {L.}~\bibnamefont
  {Embon}}, \bibinfo {author} {\bibfnamefont {Y.}~\bibnamefont {Anahory}},
  \bibinfo {author} {\bibfnamefont {Z.~L.}\ \bibnamefont {Jelic}}, \bibinfo
  {author} {\bibfnamefont {E.~O.}\ \bibnamefont {Lachman}}, \bibinfo {author}
  {\bibfnamefont {Y.}~\bibnamefont {Myasoedov}}, \bibinfo {author}
  {\bibfnamefont {M.~E.}\ \bibnamefont {Huber}}, \bibinfo {author}
  {\bibfnamefont {G.~P.}\ \bibnamefont {Mikitik}}, \bibinfo {author}
  {\bibfnamefont {A.~V.}\ \bibnamefont {Silhanek}}, \bibinfo {author}
  {\bibfnamefont {M.~V.}\ \bibnamefont {Milosevic}}, \bibinfo {author}
  {\bibfnamefont {A.}~\bibnamefont {Gurevich}}, \ and\ \bibinfo {author}
  {\bibfnamefont {E.}~\bibnamefont {Zeldov}},\ }\bibfield  {title} {\enquote
  {\bibinfo {title} {Imaging of super-fast dynamics and flow instabilities of
  superconducting vortices},}\ }\href {\doibase 10.1038/s41467-017-00089-3}
  {\bibfield  {journal} {\bibinfo  {journal} {Nat. Commun.}\ }\textbf {\bibinfo
  {volume} {8}},\ \bibinfo {pages} {85} (\bibinfo {year} {2017})}\BibitemShut
  {NoStop}%
\bibitem [{\citenamefont {Rouco}\ \emph {et~al.}(2018)\citenamefont {Rouco},
  \citenamefont {Massarotti}, \citenamefont {Stornaiuolo}, \citenamefont
  {Papari}, \citenamefont {Obradors}, \citenamefont {Puig}, \citenamefont
  {Tafuri},\ and\ \citenamefont {Palau}}]{Rou18mat}%
  \BibitemOpen
  \bibfield  {author} {\bibinfo {author} {\bibfnamefont {V.}~\bibnamefont
  {Rouco}}, \bibinfo {author} {\bibfnamefont {M.}~\bibnamefont {Massarotti}},
  \bibinfo {author} {\bibfnamefont {D.}~\bibnamefont {Stornaiuolo}}, \bibinfo
  {author} {\bibfnamefont {G.~P.}\ \bibnamefont {Papari}}, \bibinfo {author}
  {\bibfnamefont {X.}~\bibnamefont {Obradors}}, \bibinfo {author}
  {\bibfnamefont {T.}~\bibnamefont {Puig}}, \bibinfo {author} {\bibfnamefont
  {F.}~\bibnamefont {Tafuri}}, \ and\ \bibinfo {author} {\bibfnamefont
  {A.}~\bibnamefont {Palau}},\ }\bibfield  {title} {\enquote {\bibinfo {title}
  {Vortex lattice instabilities in {YBa$_2$Cu$_3$O$_{7-x}$} nanowires},}\
  }\href {\doibase 10.3390/ma11020211} {\bibfield  {journal} {\bibinfo
  {journal} {Materials}\ }\textbf {\bibinfo {volume} {11}},\ \bibinfo {pages}
  {211} (\bibinfo {year} {2018})}\BibitemShut {NoStop}%
\bibitem [{\citenamefont {Dobrovolskiy}\ \emph
  {et~al.}(2019{\natexlab{a}})\citenamefont {Dobrovolskiy}, \citenamefont
  {Bevz}, \citenamefont {Begun}, \citenamefont {Sachser}, \citenamefont
  {Vovk},\ and\ \citenamefont {Huth}}]{Dob19pra}%
  \BibitemOpen
  \bibfield  {author} {\bibinfo {author} {\bibfnamefont {O.~V.}\ \bibnamefont
  {Dobrovolskiy}}, \bibinfo {author} {\bibfnamefont {V.~M.}\ \bibnamefont
  {Bevz}}, \bibinfo {author} {\bibfnamefont {E.}~\bibnamefont {Begun}},
  \bibinfo {author} {\bibfnamefont {R.}~\bibnamefont {Sachser}}, \bibinfo
  {author} {\bibfnamefont {R.~V.}\ \bibnamefont {Vovk}}, \ and\ \bibinfo
  {author} {\bibfnamefont {M.}~\bibnamefont {Huth}},\ }\bibfield  {title}
  {\enquote {\bibinfo {title} {Fast dynamics of guided magnetic flux quanta},}\
  }\href {\doibase 10.1103/PhysRevApplied.11.054064} {\bibfield  {journal}
  {\bibinfo  {journal} {Phys. Rev. Appl.}\ }\textbf {\bibinfo {volume} {11}},\
  \bibinfo {pages} {054064} (\bibinfo {year} {2019}{\natexlab{a}})}\BibitemShut
  {NoStop}%
\bibitem [{\citenamefont {Leo}\ \emph {et~al.}(2020)\citenamefont {Leo},
  \citenamefont {Nigro}, \citenamefont {Braccini}, \citenamefont {Sylva},
  \citenamefont {Provino}, \citenamefont {Galluzzi}, \citenamefont
  {Polichetti}, \citenamefont {Ferdeghini}, \citenamefont {Putti},\ and\
  \citenamefont {Grimaldi}}]{Leo20sst}%
  \BibitemOpen
  \bibfield  {author} {\bibinfo {author} {\bibfnamefont {A.}~\bibnamefont
  {Leo}}, \bibinfo {author} {\bibfnamefont {A.}~\bibnamefont {Nigro}}, \bibinfo
  {author} {\bibfnamefont {V.}~\bibnamefont {Braccini}}, \bibinfo {author}
  {\bibfnamefont {G.}~\bibnamefont {Sylva}}, \bibinfo {author} {\bibfnamefont
  {A.}~\bibnamefont {Provino}}, \bibinfo {author} {\bibfnamefont
  {A.}~\bibnamefont {Galluzzi}}, \bibinfo {author} {\bibfnamefont
  {M.}~\bibnamefont {Polichetti}}, \bibinfo {author} {\bibfnamefont
  {C.}~\bibnamefont {Ferdeghini}}, \bibinfo {author} {\bibfnamefont
  {M.}~\bibnamefont {Putti}}, \ and\ \bibinfo {author} {\bibfnamefont
  {G.}~\bibnamefont {Grimaldi}},\ }\bibfield  {title} {\enquote {\bibinfo
  {title} {Flux flow instability as a probe for quasiparticle energy relaxation
  time in {Fe}-chalcogenides},}\ }\href {\doibase 10.1088/1361-6668/abaec1}
  {\bibfield  {journal} {\bibinfo  {journal} {Supercond. Sci. Technol.}\
  }\textbf {\bibinfo {volume} {33}},\ \bibinfo {pages} {104005} (\bibinfo
  {year} {2020})}\BibitemShut {NoStop}%
\bibitem [{\citenamefont {Dobrovolskiy}\ \emph
  {et~al.}(2020{\natexlab{a}})\citenamefont {Dobrovolskiy}, \citenamefont
  {Vodolazov}, \citenamefont {Porrati}, \citenamefont {Sachser}, \citenamefont
  {Bevz}, \citenamefont {Mikhailov}, \citenamefont {Chumak},\ and\
  \citenamefont {Huth}}]{Dob20nac}%
  \BibitemOpen
  \bibfield  {author} {\bibinfo {author} {\bibfnamefont {O.~V.}\ \bibnamefont
  {Dobrovolskiy}}, \bibinfo {author} {\bibfnamefont {D.~Yu}\ \bibnamefont
  {Vodolazov}}, \bibinfo {author} {\bibfnamefont {F.}~\bibnamefont {Porrati}},
  \bibinfo {author} {\bibfnamefont {R.}~\bibnamefont {Sachser}}, \bibinfo
  {author} {\bibfnamefont {V.~M.}\ \bibnamefont {Bevz}}, \bibinfo {author}
  {\bibfnamefont {M.~Yu}\ \bibnamefont {Mikhailov}}, \bibinfo {author}
  {\bibfnamefont {A.~V.}\ \bibnamefont {Chumak}}, \ and\ \bibinfo {author}
  {\bibfnamefont {M.}~\bibnamefont {Huth}},\ }\bibfield  {title} {\enquote
  {\bibinfo {title} {Ultra-fast vortex motion in a direct-write {Nb-C}
  superconductor},}\ }\href {\doibase 10.1038/s41467-020-16987-y} {\bibfield
  {journal} {\bibinfo  {journal} {Nat. Commun.}\ }\textbf {\bibinfo {volume}
  {11}},\ \bibinfo {pages} {3291} (\bibinfo {year}
  {2020}{\natexlab{a}})}\BibitemShut {NoStop}%
\bibitem [{\citenamefont {Ustavschikov}\ \emph {et~al.}(2020)\citenamefont
  {Ustavschikov}, \citenamefont {Levichev}, \citenamefont {Pashenkin},
  \citenamefont {Klushin},\ and\ \citenamefont {Vodolazov}}]{Ust20sst}%
  \BibitemOpen
  \bibfield  {author} {\bibinfo {author} {\bibfnamefont {S.~S.}\ \bibnamefont
  {Ustavschikov}}, \bibinfo {author} {\bibfnamefont {M.~Yu.}\ \bibnamefont
  {Levichev}}, \bibinfo {author} {\bibfnamefont {I.~Yu.}\ \bibnamefont
  {Pashenkin}}, \bibinfo {author} {\bibfnamefont {A.~M.}\ \bibnamefont
  {Klushin}}, \ and\ \bibinfo {author} {\bibfnamefont {D.~Yu.}\ \bibnamefont
  {Vodolazov}},\ }\bibfield  {title} {\enquote {\bibinfo {title} {Approaching
  depairing current in dirty thin superconducting strip covered by low
  resistive normal metal},}\ }\href {\doibase 10.1088/1361-6668/abc2ad}
  {\bibfield  {journal} {\bibinfo  {journal} {Supercond. Sci. Technol.}\
  }\textbf {\bibinfo {volume} {34}},\ \bibinfo {pages} {015004} (\bibinfo
  {year} {2020})}\BibitemShut {NoStop}%
\bibitem [{\citenamefont {Vodolazov}(2019)}]{Vod19sst}%
  \BibitemOpen
  \bibfield  {author} {\bibinfo {author} {\bibfnamefont {D.~Yu.}\ \bibnamefont
  {Vodolazov}},\ }\bibfield  {title} {\enquote {\bibinfo {title} {Flux-flow
  instability in a strongly disordered superconducting strip with an edge
  barrier for vortex entry},}\ }\href {\doibase 10.1088/1361-6668/ab4168}
  {\bibfield  {journal} {\bibinfo  {journal} {Supercond. Sci. Technol.}\
  }\textbf {\bibinfo {volume} {32}},\ \bibinfo {pages} {115013} (\bibinfo
  {year} {2019})}\BibitemShut {NoStop}%
\bibitem [{\citenamefont {Bezuglyj}\ \emph {et~al.}(2019)\citenamefont
  {Bezuglyj}, \citenamefont {Shklovskij}, \citenamefont {Vovk}, \citenamefont
  {Bevz}, \citenamefont {Huth},\ and\ \citenamefont {Dobrovolskiy}}]{Bez19prb}%
  \BibitemOpen
  \bibfield  {author} {\bibinfo {author} {\bibfnamefont {Alexei~I.}\
  \bibnamefont {Bezuglyj}}, \bibinfo {author} {\bibfnamefont {Valerij~A.}\
  \bibnamefont {Shklovskij}}, \bibinfo {author} {\bibfnamefont {Ruslan~V.}\
  \bibnamefont {Vovk}}, \bibinfo {author} {\bibfnamefont {Volodymyr~M.}\
  \bibnamefont {Bevz}}, \bibinfo {author} {\bibfnamefont {Michael}\
  \bibnamefont {Huth}}, \ and\ \bibinfo {author} {\bibfnamefont {Oleksandr~V.}\
  \bibnamefont {Dobrovolskiy}},\ }\bibfield  {title} {\enquote {\bibinfo
  {title} {Local flux-flow instability in superconducting films near
  {${T}_{c}$}},}\ }\href {\doibase 10.1103/PhysRevB.99.174518} {\bibfield
  {journal} {\bibinfo  {journal} {Phys. Rev. B}\ }\textbf {\bibinfo {volume}
  {99}},\ \bibinfo {pages} {174518} (\bibinfo {year} {2019})}\BibitemShut
  {NoStop}%
\bibitem [{\citenamefont {Kogan}\ and\ \citenamefont
  {Prozorov}(2020)}]{Kog20prb}%
  \BibitemOpen
  \bibfield  {author} {\bibinfo {author} {\bibfnamefont {V.~G.}\ \bibnamefont
  {Kogan}}\ and\ \bibinfo {author} {\bibfnamefont {R.}~\bibnamefont
  {Prozorov}},\ }\bibfield  {title} {\enquote {\bibinfo {title} {Interaction
  between moving {Abrikosov} vortices in {type-II} superconductors},}\ }\href
  {\doibase 10.1103/PhysRevB.102.024506} {\bibfield  {journal} {\bibinfo
  {journal} {Phys. Rev. B}\ }\textbf {\bibinfo {volume} {102}},\ \bibinfo
  {pages} {024506} (\bibinfo {year} {2020})}\BibitemShut {NoStop}%
\bibitem [{\citenamefont {Pathirana}\ and\ \citenamefont
  {Gurevich}(2020)}]{Pat20prb}%
  \BibitemOpen
  \bibfield  {author} {\bibinfo {author} {\bibfnamefont {W.~P. M.~R.}\
  \bibnamefont {Pathirana}}\ and\ \bibinfo {author} {\bibfnamefont
  {A.}~\bibnamefont {Gurevich}},\ }\bibfield  {title} {\enquote {\bibinfo
  {title} {Nonlinear dynamics and dissipation of a curvilinear vortex driven by
  a strong time-dependent {Meissner} current},}\ }\href {\doibase
  10.1103/PhysRevB.101.064504} {\bibfield  {journal} {\bibinfo  {journal}
  {Phys. Rev. B}\ }\textbf {\bibinfo {volume} {101}},\ \bibinfo {pages}
  {064504} (\bibinfo {year} {2020})}\BibitemShut {NoStop}%
\bibitem [{\citenamefont {Kogan}\ and\ \citenamefont
  {Nakagawa}(2021)}]{Kog21prb}%
  \BibitemOpen
  \bibfield  {author} {\bibinfo {author} {\bibfnamefont {V.~G.}\ \bibnamefont
  {Kogan}}\ and\ \bibinfo {author} {\bibfnamefont {N.}~\bibnamefont
  {Nakagawa}},\ }\bibfield  {title} {\enquote {\bibinfo {title} {Current
  distributions by moving vortices in superconductors},}\ }\href {\doibase
  10.1103/PhysRevB.103.134511} {\bibfield  {journal} {\bibinfo  {journal}
  {Phys. Rev. B}\ }\textbf {\bibinfo {volume} {103}},\ \bibinfo {pages}
  {134511} (\bibinfo {year} {2021})}\BibitemShut {NoStop}%
\bibitem [{\citenamefont {Pathirana}\ and\ \citenamefont
  {Gurevich}(2021)}]{Pat21prb}%
  \BibitemOpen
  \bibfield  {author} {\bibinfo {author} {\bibfnamefont {W.~P. M.~R.}\
  \bibnamefont {Pathirana}}\ and\ \bibinfo {author} {\bibfnamefont
  {A.}~\bibnamefont {Gurevich}},\ }\bibfield  {title} {\enquote {\bibinfo
  {title} {Effect of random pinning on nonlinear dynamics and dissipation of a
  vortex driven by a strong microwave current},}\ }\href {\doibase
  10.1103/PhysRevB.103.184518} {\bibfield  {journal} {\bibinfo  {journal}
  {Phys. Rev. B}\ }\textbf {\bibinfo {volume} {103}},\ \bibinfo {pages}
  {184518} (\bibinfo {year} {2021})}\BibitemShut {NoStop}%
\bibitem [{\citenamefont {Ivlev}\ \emph {et~al.}(1999)\citenamefont {Ivlev},
  \citenamefont {Mej\'{\i}a-Rosales},\ and\ \citenamefont
  {Kunchur}}]{Ivl99prb}%
  \BibitemOpen
  \bibfield  {author} {\bibinfo {author} {\bibfnamefont {B.~I.}\ \bibnamefont
  {Ivlev}}, \bibinfo {author} {\bibfnamefont {S.}~\bibnamefont
  {Mej\'{\i}a-Rosales}}, \ and\ \bibinfo {author} {\bibfnamefont {M.~N.}\
  \bibnamefont {Kunchur}},\ }\bibfield  {title} {\enquote {\bibinfo {title}
  {Cherenkov resonances in vortex dissipation in superconductors},}\ }\href
  {\doibase 10.1103/PhysRevB.60.12419} {\bibfield  {journal} {\bibinfo
  {journal} {Phys. Rev. B}\ }\textbf {\bibinfo {volume} {60}},\ \bibinfo
  {pages} {12419--12423} (\bibinfo {year} {1999})}\BibitemShut {NoStop}%
\bibitem [{\citenamefont {Bulaevskii}\ and\ \citenamefont
  {Chudnovsky}(2005)}]{Bul05prb}%
  \BibitemOpen
  \bibfield  {author} {\bibinfo {author} {\bibfnamefont {L.~N.}\ \bibnamefont
  {Bulaevskii}}\ and\ \bibinfo {author} {\bibfnamefont {E.~M.}\ \bibnamefont
  {Chudnovsky}},\ }\bibfield  {title} {\enquote {\bibinfo {title} {Sound
  generation by the vortex flow in type-{II} superconductors},}\ }\href
  {\doibase 10.1103/PhysRevB.72.094518} {\bibfield  {journal} {\bibinfo
  {journal} {Phys. Rev. B}\ }\textbf {\bibinfo {volume} {72}},\ \bibinfo
  {pages} {094518} (\bibinfo {year} {2005})}\BibitemShut {NoStop}%
\bibitem [{\citenamefont {Bespalov}\ \emph {et~al.}(2014)\citenamefont
  {Bespalov}, \citenamefont {Mel'nikov},\ and\ \citenamefont
  {Buzdin}}]{Bes14prb}%
  \BibitemOpen
  \bibfield  {author} {\bibinfo {author} {\bibfnamefont {A.~A.}\ \bibnamefont
  {Bespalov}}, \bibinfo {author} {\bibfnamefont {A.~S.}\ \bibnamefont
  {Mel'nikov}}, \ and\ \bibinfo {author} {\bibfnamefont {A.~I.}\ \bibnamefont
  {Buzdin}},\ }\bibfield  {title} {\enquote {\bibinfo {title} {Magnon radiation
  by moving {Abrikosov} vortices in ferromagnetic superconductors and
  superconductor-ferromagnet multilayers},}\ }\href {\doibase
  10.1103/PhysRevB.89.054516} {\bibfield  {journal} {\bibinfo  {journal} {Phys.
  Rev. B}\ }\textbf {\bibinfo {volume} {89}},\ \bibinfo {pages} {054516}
  (\bibinfo {year} {2014})}\BibitemShut {NoStop}%
\bibitem [{\citenamefont {Dobrovolskiy}\ \emph {et~al.}(2021)\citenamefont
  {Dobrovolskiy}, \citenamefont {Wang}, \citenamefont {Vodolazov},
  \citenamefont {Budinska}, \citenamefont {Sachser}, \citenamefont {Chumak},
  \citenamefont {Huth},\ and\ \citenamefont {Buzdin}}]{Dob21arx}%
  \BibitemOpen
  \bibfield  {author} {\bibinfo {author} {\bibfnamefont {O.~V.}\ \bibnamefont
  {Dobrovolskiy}}, \bibinfo {author} {\bibfnamefont {Q.}~\bibnamefont {Wang}},
  \bibinfo {author} {\bibfnamefont {D.~Yu.}\ \bibnamefont {Vodolazov}},
  \bibinfo {author} {\bibfnamefont {B.}~\bibnamefont {Budinska}}, \bibinfo
  {author} {\bibfnamefont {R.}~\bibnamefont {Sachser}}, \bibinfo {author}
  {\bibfnamefont {A.V.}\ \bibnamefont {Chumak}}, \bibinfo {author}
  {\bibfnamefont {M.}~\bibnamefont {Huth}}, \ and\ \bibinfo {author}
  {\bibfnamefont {A.~I.}\ \bibnamefont {Buzdin}},\ }\bibfield  {title}
  {\enquote {\bibinfo {title} {Cherenkov radiation of spin waves by ultra-fast
  moving magnetic flux quanta},}\ }\href@noop {} {\bibfield  {journal}
  {\bibinfo  {journal} {arXiv:2103.10156}\ } (\bibinfo {year}
  {2021})}\BibitemShut {NoStop}%
\bibitem [{\citenamefont {Korneeva}\ \emph {et~al.}(2018)\citenamefont
  {Korneeva}, \citenamefont {Vodolazov}, \citenamefont {Semenov}, \citenamefont
  {Florya}, \citenamefont {Simonov}, \citenamefont {Baeva}, \citenamefont
  {Korneev}, \citenamefont {Goltsman},\ and\ \citenamefont
  {Klapwijk}}]{Kor18pra}%
  \BibitemOpen
  \bibfield  {author} {\bibinfo {author} {\bibfnamefont {Yu.~P.}\ \bibnamefont
  {Korneeva}}, \bibinfo {author} {\bibfnamefont {D.~Yu.}\ \bibnamefont
  {Vodolazov}}, \bibinfo {author} {\bibfnamefont {A.~V.}\ \bibnamefont
  {Semenov}}, \bibinfo {author} {\bibfnamefont {I.~N.}\ \bibnamefont {Florya}},
  \bibinfo {author} {\bibfnamefont {N.}~\bibnamefont {Simonov}}, \bibinfo
  {author} {\bibfnamefont {E.}~\bibnamefont {Baeva}}, \bibinfo {author}
  {\bibfnamefont {A.~A.}\ \bibnamefont {Korneev}}, \bibinfo {author}
  {\bibfnamefont {G.~N.}\ \bibnamefont {Goltsman}}, \ and\ \bibinfo {author}
  {\bibfnamefont {T.~M.}\ \bibnamefont {Klapwijk}},\ }\bibfield  {title}
  {\enquote {\bibinfo {title} {Optical single-photon detection in
  micrometer-scale {NbN} bridges},}\ }\href {\doibase
  10.1103/PhysRevApplied.9.064037} {\bibfield  {journal} {\bibinfo  {journal}
  {Phys. Rev. Appl.}\ }\textbf {\bibinfo {volume} {9}},\ \bibinfo {pages}
  {064037} (\bibinfo {year} {2018})}\BibitemShut {NoStop}%
\bibitem [{\citenamefont {Korneeva}\ \emph {et~al.}(2020)\citenamefont
  {Korneeva}, \citenamefont {Manova}, \citenamefont {Florya}, \citenamefont
  {Mikhailov}, \citenamefont {Dobrovolskiy}, \citenamefont {Korneev},\ and\
  \citenamefont {Vodolazov}}]{Kor20pra}%
  \BibitemOpen
  \bibfield  {author} {\bibinfo {author} {\bibfnamefont {Yu.~P.}\ \bibnamefont
  {Korneeva}}, \bibinfo {author} {\bibfnamefont {N.N.}\ \bibnamefont {Manova}},
  \bibinfo {author} {\bibfnamefont {I.N.}\ \bibnamefont {Florya}}, \bibinfo
  {author} {\bibfnamefont {M.~Yu.}\ \bibnamefont {Mikhailov}}, \bibinfo
  {author} {\bibfnamefont {O.V.}\ \bibnamefont {Dobrovolskiy}}, \bibinfo
  {author} {\bibfnamefont {A.A.}\ \bibnamefont {Korneev}}, \ and\ \bibinfo
  {author} {\bibfnamefont {D.~Yu.}\ \bibnamefont {Vodolazov}},\ }\bibfield
  {title} {\enquote {\bibinfo {title} {Different single-photon response of wide
  and narrow superconducting
  {${\mathrm{Mo}}_{x}{\mathrm{Si}}_{1\ensuremath{-}x}$} strips},}\ }\href
  {\doibase 10.1103/PhysRevApplied.13.024011} {\bibfield  {journal} {\bibinfo
  {journal} {Phys. Rev. Appl.}\ }\textbf {\bibinfo {volume} {13}},\ \bibinfo
  {pages} {024011} (\bibinfo {year} {2020})}\BibitemShut {NoStop}%
\bibitem [{\citenamefont {Charaev}\ \emph {et~al.}(2020)\citenamefont
  {Charaev}, \citenamefont {Morimoto}, \citenamefont {Dane}, \citenamefont
  {Agarwal}, \citenamefont {Colangelo},\ and\ \citenamefont
  {Berggren}}]{Cha20apl}%
  \BibitemOpen
  \bibfield  {author} {\bibinfo {author} {\bibfnamefont {I.}~\bibnamefont
  {Charaev}}, \bibinfo {author} {\bibfnamefont {Y.}~\bibnamefont {Morimoto}},
  \bibinfo {author} {\bibfnamefont {A.}~\bibnamefont {Dane}}, \bibinfo {author}
  {\bibfnamefont {A.}~\bibnamefont {Agarwal}}, \bibinfo {author} {\bibfnamefont
  {M.}~\bibnamefont {Colangelo}}, \ and\ \bibinfo {author} {\bibfnamefont
  {K.~K.}\ \bibnamefont {Berggren}},\ }\bibfield  {title} {\enquote {\bibinfo
  {title} {Large-area microwire {MoSi} single-photon detectors at 1550 nm
  wavelength},}\ }\href {\doibase 10.1063/5.0005439} {\bibfield  {journal}
  {\bibinfo  {journal} {Appl. Phys. Lett.}\ }\textbf {\bibinfo {volume}
  {116}},\ \bibinfo {pages} {242603} (\bibinfo {year} {2020})}\BibitemShut
  {NoStop}%
\bibitem [{\citenamefont {Chiles}\ \emph {et~al.}(2020)\citenamefont {Chiles},
  \citenamefont {Buckley}, \citenamefont {Lita}, \citenamefont {Verma},
  \citenamefont {Allmaras}, \citenamefont {Korzh}, \citenamefont {Shaw},
  \citenamefont {Shainline}, \citenamefont {Mirin},\ and\ \citenamefont
  {Nam}}]{Chi20apl}%
  \BibitemOpen
  \bibfield  {author} {\bibinfo {author} {\bibfnamefont {J.}~\bibnamefont
  {Chiles}}, \bibinfo {author} {\bibfnamefont {S.~M.}\ \bibnamefont {Buckley}},
  \bibinfo {author} {\bibfnamefont {A.}~\bibnamefont {Lita}}, \bibinfo {author}
  {\bibfnamefont {V.~B.}\ \bibnamefont {Verma}}, \bibinfo {author}
  {\bibfnamefont {J.}~\bibnamefont {Allmaras}}, \bibinfo {author}
  {\bibfnamefont {B.}~\bibnamefont {Korzh}}, \bibinfo {author} {\bibfnamefont
  {M.~D.}\ \bibnamefont {Shaw}}, \bibinfo {author} {\bibfnamefont {J.~M.}\
  \bibnamefont {Shainline}}, \bibinfo {author} {\bibfnamefont {R.~P.}\
  \bibnamefont {Mirin}}, \ and\ \bibinfo {author} {\bibfnamefont {S.~W.}\
  \bibnamefont {Nam}},\ }\bibfield  {title} {\enquote {\bibinfo {title}
  {Superconducting microwire detectors based on {WSi} with single-photon
  sensitivity in the near-infrared},}\ }\href {\doibase 10.1063/5.0006221}
  {\bibfield  {journal} {\bibinfo  {journal} {Appl. Phys. Lett.}\ }\textbf
  {\bibinfo {volume} {116}},\ \bibinfo {pages} {242602} (\bibinfo {year}
  {2020})}\BibitemShut {NoStop}%
\bibitem [{\citenamefont {Vodolazov}(2017)}]{Vod17pra}%
  \BibitemOpen
  \bibfield  {author} {\bibinfo {author} {\bibfnamefont {D.~Yu.}\ \bibnamefont
  {Vodolazov}},\ }\bibfield  {title} {\enquote {\bibinfo {title} {Single-photon
  detection by a dirty current-carrying superconducting strip based on the
  kinetic-equation approach},}\ }\href {\doibase
  10.1103/PhysRevApplied.7.034014} {\bibfield  {journal} {\bibinfo  {journal}
  {Phys. Rev. Appl.}\ }\textbf {\bibinfo {volume} {7}},\ \bibinfo {pages}
  {034014} (\bibinfo {year} {2017})}\BibitemShut {NoStop}%
\bibitem [{\citenamefont {Caputo}\ \emph {et~al.}(2017)\citenamefont {Caputo},
  \citenamefont {Cirillo},\ and\ \citenamefont {Attanasio}}]{Cap17apl}%
  \BibitemOpen
  \bibfield  {author} {\bibinfo {author} {\bibfnamefont {M.}~\bibnamefont
  {Caputo}}, \bibinfo {author} {\bibfnamefont {C.}~\bibnamefont {Cirillo}}, \
  and\ \bibinfo {author} {\bibfnamefont {C.}~\bibnamefont {Attanasio}},\
  }\bibfield  {title} {\enquote {\bibinfo {title} {{NbRe} as candidate material
  for fast single photon detection},}\ }\href {\doibase 10.1063/1.4997675}
  {\bibfield  {journal} {\bibinfo  {journal} {Appl. Phys. Lett.}\ }\textbf
  {\bibinfo {volume} {111}},\ \bibinfo {pages} {192601} (\bibinfo {year}
  {2017})}\BibitemShut {NoStop}%
\bibitem [{\citenamefont {Hofer}\ and\ \citenamefont
  {Haberkorn}(2021)}]{Hof21tsf}%
  \BibitemOpen
  \bibfield  {author} {\bibinfo {author} {\bibfnamefont {J.~A.}\ \bibnamefont
  {Hofer}}\ and\ \bibinfo {author} {\bibfnamefont {N.}~\bibnamefont
  {Haberkorn}},\ }\bibfield  {title} {\enquote {\bibinfo {title} {Flux flow
  velocity instability and quasiparticle relaxation time in nanocrystalline
  $\beta$-{W} thin films},}\ }\href
  {https://www.sciencedirect.com/science/article/pii/S0040609021001735}
  {\bibfield  {journal} {\bibinfo  {journal} {Thin Sol. Films}\ }\textbf
  {\bibinfo {volume} {730}},\ \bibinfo {pages} {138690} (\bibinfo {year}
  {2021})}\BibitemShut {NoStop}%
\bibitem [{\citenamefont {Liu}\ \emph {et~al.}(2021)\citenamefont {Liu},
  \citenamefont {Luo}, \citenamefont {Zhang}, \citenamefont {Hou},\ and\
  \citenamefont {Wang}}]{Liu21sst}%
  \BibitemOpen
  \bibfield  {author} {\bibinfo {author} {\bibfnamefont {Z.}~\bibnamefont
  {Liu}}, \bibinfo {author} {\bibfnamefont {B.}~\bibnamefont {Luo}}, \bibinfo
  {author} {\bibfnamefont {L.}~\bibnamefont {Zhang}}, \bibinfo {author}
  {\bibfnamefont {B.}~\bibnamefont {Hou}}, \ and\ \bibinfo {author}
  {\bibfnamefont {D.}~\bibnamefont {Wang}},\ }\bibfield  {title} {\enquote
  {\bibinfo {title} {Vortex dynamics in amorphous {MoSi} superconducting thin
  films},}\ }\href {http://iopscience.iop.org/article/10.1088/1361-6668/ac2eb0}
  {\bibfield  {journal} {\bibinfo  {journal} {Supercond. Sci. Technol.}\ }
  (\bibinfo {year} {2021})}\BibitemShut {NoStop}%
\bibitem [{\citenamefont {Cirillo}\ \emph {et~al.}(2021)\citenamefont
  {Cirillo}, \citenamefont {Granata}, \citenamefont {Spuri}, \citenamefont
  {Di~Bernardo},\ and\ \citenamefont {Attanasio}}]{Cir21prm}%
  \BibitemOpen
  \bibfield  {author} {\bibinfo {author} {\bibfnamefont {C.}~\bibnamefont
  {Cirillo}}, \bibinfo {author} {\bibfnamefont {V.}~\bibnamefont {Granata}},
  \bibinfo {author} {\bibfnamefont {A.}~\bibnamefont {Spuri}}, \bibinfo
  {author} {\bibfnamefont {A.}~\bibnamefont {Di~Bernardo}}, \ and\ \bibinfo
  {author} {\bibfnamefont {C.}~\bibnamefont {Attanasio}},\ }\bibfield  {title}
  {\enquote {\bibinfo {title} {{NbReN: A} disordered superconductor in thin
  film form for potential application as superconducting nanowire single photon
  detector},}\ }\href {\doibase 10.1103/PhysRevMaterials.5.085004} {\bibfield
  {journal} {\bibinfo  {journal} {Phys. Rev. Mater.}\ }\textbf {\bibinfo
  {volume} {5}},\ \bibinfo {pages} {085004} (\bibinfo {year}
  {2021})}\BibitemShut {NoStop}%
\bibitem [{\citenamefont {Larkin}\ and\ \citenamefont
  {Ovchinnikov}(1975)}]{Lar75etp}%
  \BibitemOpen
  \bibfield  {author} {\bibinfo {author} {\bibfnamefont {A.~I.}\ \bibnamefont
  {Larkin}}\ and\ \bibinfo {author} {\bibfnamefont {Yu.~N.}\ \bibnamefont
  {Ovchinnikov}},\ }\bibfield  {title} {\enquote {\bibinfo {title} {Nonlinear
  conductivity of superconductors in the mixed state},}\ }\href
  {http://jetp.ras.ru/cgi-bin/e/index/e/41/5/p960?a=list} {\bibfield  {journal}
  {\bibinfo  {journal} {J. Exp. Theor. Phys.}\ }\textbf {\bibinfo {volume}
  {41}},\ \bibinfo {pages} {960} (\bibinfo {year} {1975})}\BibitemShut
  {NoStop}%
\bibitem [{\citenamefont {Larkin}\ and\ \citenamefont
  {Ovchinnikov}(1986)}]{Lar86inb}%
  \BibitemOpen
  \bibfield  {author} {\bibinfo {author} {\bibfnamefont {A.~I.}\ \bibnamefont
  {Larkin}}\ and\ \bibinfo {author} {\bibfnamefont {Y.~N.}\ \bibnamefont
  {Ovchinnikov}},\ }\enquote {\bibinfo {title} {Nonequilibrium
  superconductivity},}\ \ (\bibinfo  {publisher} {Elsevier, Amsterdam},\
  \bibinfo {year} {1986})\ p.\ \bibinfo {pages} {493}\BibitemShut {NoStop}%
\bibitem [{\citenamefont {Bezuglyj}\ and\ \citenamefont
  {Shklovskij}(1992)}]{Bez92pcs}%
  \BibitemOpen
  \bibfield  {author} {\bibinfo {author} {\bibfnamefont {A.I.}\ \bibnamefont
  {Bezuglyj}}\ and\ \bibinfo {author} {\bibfnamefont {V.A.}\ \bibnamefont
  {Shklovskij}},\ }\bibfield  {title} {\enquote {\bibinfo {title} {Effect of
  self-heating on flux flow instability in a superconductor near {$T_c$}},}\
  }\href {\doibase 10.1016/0921-4534(92)90165-9} {\bibfield  {journal}
  {\bibinfo  {journal} {Physica C}\ }\textbf {\bibinfo {volume} {202}},\
  \bibinfo {pages} {234} (\bibinfo {year} {1992})}\BibitemShut {NoStop}%
\bibitem [{\citenamefont {Silhanek}\ \emph {et~al.}(2012)\citenamefont
  {Silhanek}, \citenamefont {Leo}, \citenamefont {Grimaldi}, \citenamefont
  {Berdiyorov}, \citenamefont {Milosevic}, \citenamefont {Nigro}, \citenamefont
  {Pace}, \citenamefont {Verellen}, \citenamefont {Gillijns}, \citenamefont
  {Metlushko}, \citenamefont {Ili\'c}, \citenamefont {Zhu},\ and\ \citenamefont
  {Moshchalkov}}]{Sil12njp}%
  \BibitemOpen
  \bibfield  {author} {\bibinfo {author} {\bibfnamefont {A.~V.}\ \bibnamefont
  {Silhanek}}, \bibinfo {author} {\bibfnamefont {A.}~\bibnamefont {Leo}},
  \bibinfo {author} {\bibfnamefont {G.}~\bibnamefont {Grimaldi}}, \bibinfo
  {author} {\bibfnamefont {G.~R.}\ \bibnamefont {Berdiyorov}}, \bibinfo
  {author} {\bibfnamefont {M.~V}\ \bibnamefont {Milosevic}}, \bibinfo {author}
  {\bibfnamefont {A.}~\bibnamefont {Nigro}}, \bibinfo {author} {\bibfnamefont
  {S.}~\bibnamefont {Pace}}, \bibinfo {author} {\bibfnamefont {N.}~\bibnamefont
  {Verellen}}, \bibinfo {author} {\bibfnamefont {W.}~\bibnamefont {Gillijns}},
  \bibinfo {author} {\bibfnamefont {V.}~\bibnamefont {Metlushko}}, \bibinfo
  {author} {\bibfnamefont {B.}~\bibnamefont {Ili\'c}}, \bibinfo {author}
  {\bibfnamefont {X.}~\bibnamefont {Zhu}}, \ and\ \bibinfo {author}
  {\bibfnamefont {V.~V.}\ \bibnamefont {Moshchalkov}},\ }\bibfield  {title}
  {\enquote {\bibinfo {title} {Influence of artificial pinning on vortex
  lattice instability in superconducting films},}\ }\href
  {http://stacks.iop.org/1367-2630/14/i=5/a=053006} {\bibfield  {journal}
  {\bibinfo  {journal} {New J. Phys.}\ }\textbf {\bibinfo {volume} {14}},\
  \bibinfo {pages} {053006} (\bibinfo {year} {2012})}\BibitemShut {NoStop}%
\bibitem [{\citenamefont {Shklovskij}\ \emph {et~al.}(2017)\citenamefont
  {Shklovskij}, \citenamefont {Nazipova},\ and\ \citenamefont
  {Dobrovolskiy}}]{Shk17prb}%
  \BibitemOpen
  \bibfield  {author} {\bibinfo {author} {\bibfnamefont {V.~A.}\ \bibnamefont
  {Shklovskij}}, \bibinfo {author} {\bibfnamefont {A.~P.}\ \bibnamefont
  {Nazipova}}, \ and\ \bibinfo {author} {\bibfnamefont {O.~V.}\ \bibnamefont
  {Dobrovolskiy}},\ }\bibfield  {title} {\enquote {\bibinfo {title} {Pinning
  effects on self-heating and flux-flow instability in superconducting films
  near ${T}_{c}$},}\ }\href {\doibase 10.1103/PhysRevB.95.184517} {\bibfield
  {journal} {\bibinfo  {journal} {Phys. Rev. B}\ }\textbf {\bibinfo {volume}
  {95}},\ \bibinfo {pages} {184517} (\bibinfo {year} {2017})}\BibitemShut
  {NoStop}%
\bibitem [{\citenamefont {Dobrovolskiy}\ \emph {et~al.}(2017)\citenamefont
  {Dobrovolskiy}, \citenamefont {Shklovskij}, \citenamefont {Hanefeld},
  \citenamefont {Z\"orb}, \citenamefont {K\"ohs},\ and\ \citenamefont
  {Huth}}]{Dob17sst}%
  \BibitemOpen
  \bibfield  {author} {\bibinfo {author} {\bibfnamefont {O.~V.}\ \bibnamefont
  {Dobrovolskiy}}, \bibinfo {author} {\bibfnamefont {V.~A.}\ \bibnamefont
  {Shklovskij}}, \bibinfo {author} {\bibfnamefont {M.}~\bibnamefont
  {Hanefeld}}, \bibinfo {author} {\bibfnamefont {M.}~\bibnamefont {Z\"orb}},
  \bibinfo {author} {\bibfnamefont {L.}~\bibnamefont {K\"ohs}}, \ and\ \bibinfo
  {author} {\bibfnamefont {M.}~\bibnamefont {Huth}},\ }\bibfield  {title}
  {\enquote {\bibinfo {title} {Pinning effects on flux flow instability in
  epitaxial {Nb} thin films},}\ }\href
  {http://stacks.iop.org/0953-2048/30/i=8/a=085002} {\bibfield  {journal}
  {\bibinfo  {journal} {Supercond. Sci. Technol.}\ }\textbf {\bibinfo {volume}
  {30}},\ \bibinfo {pages} {085002} (\bibinfo {year} {2017})}\BibitemShut
  {NoStop}%
\bibitem [{\citenamefont {Dobrovolskiy}\ \emph
  {et~al.}(2020{\natexlab{b}})\citenamefont {Dobrovolskiy}, \citenamefont
  {Gonz{\'a}lez-Ruano}, \citenamefont {Lara}, \citenamefont {Sachser},
  \citenamefont {Bevz}, \citenamefont {Shklovskij}, \citenamefont {Bezuglyj},
  \citenamefont {Vovk}, \citenamefont {Huth},\ and\ \citenamefont
  {Aliev}}]{Dob20cph}%
  \BibitemOpen
  \bibfield  {author} {\bibinfo {author} {\bibfnamefont {O.~V.}\ \bibnamefont
  {Dobrovolskiy}}, \bibinfo {author} {\bibfnamefont {C.}~\bibnamefont
  {Gonz{\'a}lez-Ruano}}, \bibinfo {author} {\bibfnamefont {A.}~\bibnamefont
  {Lara}}, \bibinfo {author} {\bibfnamefont {R.}~\bibnamefont {Sachser}},
  \bibinfo {author} {\bibfnamefont {V.~M.}\ \bibnamefont {Bevz}}, \bibinfo
  {author} {\bibfnamefont {V.~A.}\ \bibnamefont {Shklovskij}}, \bibinfo
  {author} {\bibfnamefont {A.~I.}\ \bibnamefont {Bezuglyj}}, \bibinfo {author}
  {\bibfnamefont {R.~V.}\ \bibnamefont {Vovk}}, \bibinfo {author}
  {\bibfnamefont {M.}~\bibnamefont {Huth}}, \ and\ \bibinfo {author}
  {\bibfnamefont {F.~G.}\ \bibnamefont {Aliev}},\ }\bibfield  {title} {\enquote
  {\bibinfo {title} {Moving flux quanta cool superconductors by a microwave
  breath},}\ }\href {\doibase 10.1038/s42005-020-0329-z} {\bibfield  {journal}
  {\bibinfo  {journal} {Commun. Phys.}\ }\textbf {\bibinfo {volume} {3}},\
  \bibinfo {pages} {64} (\bibinfo {year} {2020}{\natexlab{b}})}\BibitemShut
  {NoStop}%
\bibitem [{\citenamefont {Watts-Tobin}\ \emph {et~al.}(1981)\citenamefont
  {Watts-Tobin}, \citenamefont {Kr{\"a}henb{\"u}hl},\ and\ \citenamefont
  {Kramer}}]{Wat81ltp}%
  \BibitemOpen
  \bibfield  {author} {\bibinfo {author} {\bibfnamefont {R.~J.}\ \bibnamefont
  {Watts-Tobin}}, \bibinfo {author} {\bibfnamefont {Y.}~\bibnamefont
  {Kr{\"a}henb{\"u}hl}}, \ and\ \bibinfo {author} {\bibfnamefont
  {L.}~\bibnamefont {Kramer}},\ }\bibfield  {title} {\enquote {\bibinfo {title}
  {Nonequilibrium theory of dirty, current-carrying superconductors: phase-slip
  oscillators in narrow filaments near {Tc}},}\ }\href {\doibase
  10.1007/BF00117427} {\bibfield  {journal} {\bibinfo  {journal} {J. Low Temp.
  Phys.}\ }\textbf {\bibinfo {volume} {42}},\ \bibinfo {pages} {459--501}
  (\bibinfo {year} {1981})}\BibitemShut {NoStop}%
\bibitem [{\citenamefont {Korzh}\ \emph {et~al.}(2020)\citenamefont {Korzh},
  \citenamefont {Zhao}, \citenamefont {Allmaras}, \citenamefont {Frasca},
  \citenamefont {Autry}, \citenamefont {Bersin}, \citenamefont {Beyer},
  \citenamefont {Briggs}, \citenamefont {Bumble}, \citenamefont {Colangelo},
  \citenamefont {Crouch}, \citenamefont {Dane}, \citenamefont {Gerrits},
  \citenamefont {Lita}, \citenamefont {Marsili}, \citenamefont {Moody},
  \citenamefont {Pe{\~{n}}a}, \citenamefont {Ramirez}, \citenamefont {Rezac},
  \citenamefont {Sinclair}, \citenamefont {Stevens}, \citenamefont {Velasco},
  \citenamefont {Verma}, \citenamefont {Wollman}, \citenamefont {Xie},
  \citenamefont {Zhu}, \citenamefont {Hale}, \citenamefont {Spiropulu},
  \citenamefont {Silverman}, \citenamefont {Mirin}, \citenamefont {Nam},
  \citenamefont {Kozorezov}, \citenamefont {Shaw},\ and\ \citenamefont
  {Berggren}}]{Kor20nph}%
  \BibitemOpen
  \bibfield  {author} {\bibinfo {author} {\bibfnamefont {B.}~\bibnamefont
  {Korzh}}, \bibinfo {author} {\bibfnamefont {Q.-Y.}\ \bibnamefont {Zhao}},
  \bibinfo {author} {\bibfnamefont {J.~P.}\ \bibnamefont {Allmaras}}, \bibinfo
  {author} {\bibfnamefont {S.}~\bibnamefont {Frasca}}, \bibinfo {author}
  {\bibfnamefont {T.~M.}\ \bibnamefont {Autry}}, \bibinfo {author}
  {\bibfnamefont {E.~A.}\ \bibnamefont {Bersin}}, \bibinfo {author}
  {\bibfnamefont {A.~D.}\ \bibnamefont {Beyer}}, \bibinfo {author}
  {\bibfnamefont {R.~M.}\ \bibnamefont {Briggs}}, \bibinfo {author}
  {\bibfnamefont {B.}~\bibnamefont {Bumble}}, \bibinfo {author} {\bibfnamefont
  {M.}~\bibnamefont {Colangelo}}, \bibinfo {author} {\bibfnamefont {G.~M.}\
  \bibnamefont {Crouch}}, \bibinfo {author} {\bibfnamefont {A.~E.}\
  \bibnamefont {Dane}}, \bibinfo {author} {\bibfnamefont {T.}~\bibnamefont
  {Gerrits}}, \bibinfo {author} {\bibfnamefont {A.~E.}\ \bibnamefont {Lita}},
  \bibinfo {author} {\bibfnamefont {F.}~\bibnamefont {Marsili}}, \bibinfo
  {author} {\bibfnamefont {G.}~\bibnamefont {Moody}}, \bibinfo {author}
  {\bibfnamefont {C.}~\bibnamefont {Pe{\~{n}}a}}, \bibinfo {author}
  {\bibfnamefont {E.}~\bibnamefont {Ramirez}}, \bibinfo {author} {\bibfnamefont
  {J.~D.}\ \bibnamefont {Rezac}}, \bibinfo {author} {\bibfnamefont
  {N.}~\bibnamefont {Sinclair}}, \bibinfo {author} {\bibfnamefont {M.~J.}\
  \bibnamefont {Stevens}}, \bibinfo {author} {\bibfnamefont {A.~E.}\
  \bibnamefont {Velasco}}, \bibinfo {author} {\bibfnamefont {V.~B.}\
  \bibnamefont {Verma}}, \bibinfo {author} {\bibfnamefont {E.~E.}\ \bibnamefont
  {Wollman}}, \bibinfo {author} {\bibfnamefont {S.}~\bibnamefont {Xie}},
  \bibinfo {author} {\bibfnamefont {D.}~\bibnamefont {Zhu}}, \bibinfo {author}
  {\bibfnamefont {P.~D.}\ \bibnamefont {Hale}}, \bibinfo {author}
  {\bibfnamefont {M.}~\bibnamefont {Spiropulu}}, \bibinfo {author}
  {\bibfnamefont {K.~L.}\ \bibnamefont {Silverman}}, \bibinfo {author}
  {\bibfnamefont {R.~P.}\ \bibnamefont {Mirin}}, \bibinfo {author}
  {\bibfnamefont {S.~W.}\ \bibnamefont {Nam}}, \bibinfo {author} {\bibfnamefont
  {A.~G.}\ \bibnamefont {Kozorezov}}, \bibinfo {author} {\bibfnamefont {M.~D.}\
  \bibnamefont {Shaw}}, \ and\ \bibinfo {author} {\bibfnamefont {K.~K.}\
  \bibnamefont {Berggren}},\ }\bibfield  {title} {\enquote {\bibinfo {title}
  {Demonstration of sub-3 ps temporal resolution with a superconducting
  nanowire single-photon detector},}\ }\href {\doibase
  10.1038/s41566-020-0589-x} {\bibfield  {journal} {\bibinfo  {journal} {Nat.
  Photon.}\ }\textbf {\bibinfo {volume} {14}},\ \bibinfo {pages} {250--255}
  (\bibinfo {year} {2020})}\BibitemShut {NoStop}%
\bibitem [{\citenamefont {Cirillo}\ \emph {et~al.}(2020)\citenamefont
  {Cirillo}, \citenamefont {Chang}, \citenamefont {Caputo}, \citenamefont
  {Los}, \citenamefont {Dorenbos}, \citenamefont {Esmaeil~Zadeh},\ and\
  \citenamefont {Attanasio}}]{Cir20apl}%
  \BibitemOpen
  \bibfield  {author} {\bibinfo {author} {\bibfnamefont {C.}~\bibnamefont
  {Cirillo}}, \bibinfo {author} {\bibfnamefont {J.}~\bibnamefont {Chang}},
  \bibinfo {author} {\bibfnamefont {M.}~\bibnamefont {Caputo}}, \bibinfo
  {author} {\bibfnamefont {J.~W.~N.}\ \bibnamefont {Los}}, \bibinfo {author}
  {\bibfnamefont {S.}~\bibnamefont {Dorenbos}}, \bibinfo {author}
  {\bibfnamefont {I.}~\bibnamefont {Esmaeil~Zadeh}}, \ and\ \bibinfo {author}
  {\bibfnamefont {C.}~\bibnamefont {Attanasio}},\ }\bibfield  {title} {\enquote
  {\bibinfo {title} {Superconducting nanowire single photon detectors based on
  disordered {NbRe} films},}\ }\href {\doibase 10.1063/5.0021487} {\bibfield
  {journal} {\bibinfo  {journal} {Appl. Phys. Lett.}\ }\textbf {\bibinfo
  {volume} {117}},\ \bibinfo {pages} {172602} (\bibinfo {year}
  {2020})}\BibitemShut {NoStop}%
\bibitem [{\citenamefont {Samoilov}\ \emph {et~al.}(1995)\citenamefont
  {Samoilov}, \citenamefont {Konczykowski}, \citenamefont {Yeh}, \citenamefont
  {Berry},\ and\ \citenamefont {Tsuei}}]{Sam95prl}%
  \BibitemOpen
  \bibfield  {author} {\bibinfo {author} {\bibfnamefont {A.~V.}\ \bibnamefont
  {Samoilov}}, \bibinfo {author} {\bibfnamefont {M.}~\bibnamefont
  {Konczykowski}}, \bibinfo {author} {\bibfnamefont {N.~C.}\ \bibnamefont
  {Yeh}}, \bibinfo {author} {\bibfnamefont {S.}~\bibnamefont {Berry}}, \ and\
  \bibinfo {author} {\bibfnamefont {C.~C.}\ \bibnamefont {Tsuei}},\ }\bibfield
  {title} {\enquote {\bibinfo {title} {Electric-field-induced electronic
  instability in amorphous {${\mathrm{Mo}}_{3}$Si } superconducting films},}\
  }\href {\doibase 10.1103/PhysRevLett.75.4118} {\bibfield  {journal} {\bibinfo
   {journal} {Phys. Rev. Lett.}\ }\textbf {\bibinfo {volume} {75}},\ \bibinfo
  {pages} {4118--4121} (\bibinfo {year} {1995})}\BibitemShut {NoStop}%
\bibitem [{\citenamefont {Doettinger}\ \emph {et~al.}(1997)\citenamefont
  {Doettinger}, \citenamefont {Kittelberger}, \citenamefont {Huebener},\ and\
  \citenamefont {Tsuei}}]{Doe97prb}%
  \BibitemOpen
  \bibfield  {author} {\bibinfo {author} {\bibfnamefont {S.~G.}\ \bibnamefont
  {Doettinger}}, \bibinfo {author} {\bibfnamefont {S.}~\bibnamefont
  {Kittelberger}}, \bibinfo {author} {\bibfnamefont {R.~P.}\ \bibnamefont
  {Huebener}}, \ and\ \bibinfo {author} {\bibfnamefont {C.~C.}\ \bibnamefont
  {Tsuei}},\ }\bibfield  {title} {\enquote {\bibinfo {title} {Quasiparticle
  energy relaxation in the cuprate superconductors},}\ }\href {\doibase
  10.1103/PhysRevB.56.14157} {\bibfield  {journal} {\bibinfo  {journal} {Phys.
  Rev. B}\ }\textbf {\bibinfo {volume} {56}},\ \bibinfo {pages} {14157--14162}
  (\bibinfo {year} {1997})}\BibitemShut {NoStop}%
\bibitem [{\citenamefont {Lin}\ \emph {et~al.}(2013)\citenamefont {Lin},
  \citenamefont {Ayala-Valenzuela}, \citenamefont {McDonald}, \citenamefont
  {Bulaevskii}, \citenamefont {Holesinger}, \citenamefont {Ronning},
  \citenamefont {Weisse-Bernstein}, \citenamefont {Williamson}, \citenamefont
  {Mueller}, \citenamefont {Hoffbauer}, \citenamefont {Rabin},\ and\
  \citenamefont {Graf}}]{Lin13prb}%
  \BibitemOpen
  \bibfield  {author} {\bibinfo {author} {\bibfnamefont {S.-Z.}\ \bibnamefont
  {Lin}}, \bibinfo {author} {\bibfnamefont {O.}~\bibnamefont
  {Ayala-Valenzuela}}, \bibinfo {author} {\bibfnamefont {R.~D.}\ \bibnamefont
  {McDonald}}, \bibinfo {author} {\bibfnamefont {L.~N.}\ \bibnamefont
  {Bulaevskii}}, \bibinfo {author} {\bibfnamefont {T.~G.}\ \bibnamefont
  {Holesinger}}, \bibinfo {author} {\bibfnamefont {F.}~\bibnamefont {Ronning}},
  \bibinfo {author} {\bibfnamefont {N.~R.}\ \bibnamefont {Weisse-Bernstein}},
  \bibinfo {author} {\bibfnamefont {T.~L.}\ \bibnamefont {Williamson}},
  \bibinfo {author} {\bibfnamefont {A.~H.}\ \bibnamefont {Mueller}}, \bibinfo
  {author} {\bibfnamefont {M.~A.}\ \bibnamefont {Hoffbauer}}, \bibinfo {author}
  {\bibfnamefont {M.~W.}\ \bibnamefont {Rabin}}, \ and\ \bibinfo {author}
  {\bibfnamefont {M.~J.}\ \bibnamefont {Graf}},\ }\bibfield  {title} {\enquote
  {\bibinfo {title} {Characterization of the thin-film {NbN} superconductor for
  single-photon detection by transport measurements},}\ }\href {\doibase
  10.1103/PhysRevB.87.184507} {\bibfield  {journal} {\bibinfo  {journal} {Phys.
  Rev. B}\ }\textbf {\bibinfo {volume} {87}},\ \bibinfo {pages} {184507}
  (\bibinfo {year} {2013})}\BibitemShut {NoStop}%
\bibitem [{\citenamefont {Gurevich}\ and\ \citenamefont
  {Mints}(1984)}]{Gur84spu}%
  \BibitemOpen
  \bibfield  {author} {\bibinfo {author} {\bibfnamefont {A.~V.}\ \bibnamefont
  {Gurevich}}\ and\ \bibinfo {author} {\bibfnamefont {R.~G.}\ \bibnamefont
  {Mints}},\ }\href@noop {} {\bibfield  {journal} {\bibinfo  {journal} {Sov.
  Phys. Usp.}\ }\textbf {\bibinfo {volume} {27}},\ \bibinfo {pages} {19}
  (\bibinfo {year} {1984})}\BibitemShut {NoStop}%
\bibitem [{\citenamefont {Bezuglyj}\ and\ \citenamefont
  {Shklovskij}(1984)}]{Bez84ltp}%
  \BibitemOpen
  \bibfield  {author} {\bibinfo {author} {\bibfnamefont {A.~I.}\ \bibnamefont
  {Bezuglyj}}\ and\ \bibinfo {author} {\bibfnamefont {V.~A.}\ \bibnamefont
  {Shklovskij}},\ }\bibfield  {title} {\enquote {\bibinfo {title} {Thermal
  domains in inhomogeneous current-carrying superconductors. current-voltage
  characteristics and dynamics of domain formation after current jumps},}\
  }\href {\doibase 10.1007/BF00681190} {\bibfield  {journal} {\bibinfo
  {journal} {J. Low Temp. Phys.}\ }\textbf {\bibinfo {volume} {57}},\ \bibinfo
  {pages} {227--247} (\bibinfo {year} {1984})}\BibitemShut {NoStop}%
\bibitem [{\citenamefont {Buzdin}\ and\ \citenamefont
  {Daumens}(1998)}]{Buz98pcs}%
  \BibitemOpen
  \bibfield  {author} {\bibinfo {author} {\bibfnamefont {A.}~\bibnamefont
  {Buzdin}}\ and\ \bibinfo {author} {\bibfnamefont {M.}~\bibnamefont
  {Daumens}},\ }\bibfield  {title} {\enquote {\bibinfo {title} {Electromagnetic
  pinning of vortices on different types of defects},}\ }\href
  {https://www.sciencedirect.com/science/article/pii/S0921453497017292}
  {\bibfield  {journal} {\bibinfo  {journal} {Physica C}\ }\textbf {\bibinfo
  {volume} {294}},\ \bibinfo {pages} {257--269} (\bibinfo {year}
  {1998})}\BibitemShut {NoStop}%
\bibitem [{\citenamefont {Aladyshkin}\ \emph {et~al.}(2001)\citenamefont
  {Aladyshkin}, \citenamefont {Mel'nikov}, \citenamefont {Shereshevsky},\ and\
  \citenamefont {Tokman}}]{Ala01pcs}%
  \BibitemOpen
  \bibfield  {author} {\bibinfo {author} {\bibfnamefont {A.Yu.}\ \bibnamefont
  {Aladyshkin}}, \bibinfo {author} {\bibfnamefont {A.~S.}\ \bibnamefont
  {Mel'nikov}}, \bibinfo {author} {\bibfnamefont {I.~A.}\ \bibnamefont
  {Shereshevsky}}, \ and\ \bibinfo {author} {\bibfnamefont {I.~D.}\
  \bibnamefont {Tokman}},\ }\bibfield  {title} {\enquote {\bibinfo {title}
  {What is the best gate for vortex entry into {type-II} superconductor?}}\
  }\href {https://www.sciencedirect.com/science/article/pii/S092145340100288X}
  {\bibfield  {journal} {\bibinfo  {journal} {Physica C}\ }\textbf {\bibinfo
  {volume} {361}},\ \bibinfo {pages} {67--72} (\bibinfo {year}
  {2001})}\BibitemShut {NoStop}%
\bibitem [{\citenamefont {Vodolazov}\ \emph {et~al.}(2003)\citenamefont
  {Vodolazov}, \citenamefont {Maksimov},\ and\ \citenamefont
  {Brandt}}]{Vod03pcs}%
  \BibitemOpen
  \bibfield  {author} {\bibinfo {author} {\bibfnamefont {D.~Y.}\ \bibnamefont
  {Vodolazov}}, \bibinfo {author} {\bibfnamefont {I.~L.}\ \bibnamefont
  {Maksimov}}, \ and\ \bibinfo {author} {\bibfnamefont {E.~H.}\ \bibnamefont
  {Brandt}},\ }\bibfield  {title} {\enquote {\bibinfo {title} {Vortex entry
  conditions in {type-II} superconductors: {Effect} of surface defects},}\
  }\href {https://www.sciencedirect.com/science/article/pii/S0921453402018774}
  {\bibfield  {journal} {\bibinfo  {journal} {Physica C}\ }\textbf {\bibinfo
  {volume} {384}},\ \bibinfo {pages} {211--226} (\bibinfo {year}
  {2003})}\BibitemShut {NoStop}%
\bibitem [{\citenamefont {Clem}\ and\ \citenamefont
  {Berggren}(2011)}]{Cle11prb}%
  \BibitemOpen
  \bibfield  {author} {\bibinfo {author} {\bibfnamefont {J.~R.}\ \bibnamefont
  {Clem}}\ and\ \bibinfo {author} {\bibfnamefont {K.~K.}\ \bibnamefont
  {Berggren}},\ }\bibfield  {title} {\enquote {\bibinfo {title}
  {Geometry-dependent critical currents in superconducting nanocircuits},}\
  }\href {\doibase 10.1103/PhysRevB.84.174510} {\bibfield  {journal} {\bibinfo
  {journal} {Phys. Rev. B}\ }\textbf {\bibinfo {volume} {84}},\ \bibinfo
  {pages} {174510} (\bibinfo {year} {2011})}\BibitemShut {NoStop}%
\bibitem [{\citenamefont {Mikitik}(2021)}]{Mik21prb}%
  \BibitemOpen
  \bibfield  {author} {\bibinfo {author} {\bibfnamefont {G.~P.}\ \bibnamefont
  {Mikitik}},\ }\bibfield  {title} {\enquote {\bibinfo {title} {Critical
  current in thin flat superconductors with {Bean-Livingston} and geometrical
  barriers},}\ }\href {\doibase 10.1103/PhysRevB.104.094526} {\bibfield
  {journal} {\bibinfo  {journal} {Phys. Rev. B}\ }\textbf {\bibinfo {volume}
  {104}},\ \bibinfo {pages} {094526} (\bibinfo {year} {2021})}\BibitemShut
  {NoStop}%
\bibitem [{\citenamefont {Caloz}\ \emph {et~al.}(2018)\citenamefont {Caloz},
  \citenamefont {Perrenoud}, \citenamefont {Autebert}, \citenamefont {Korzh},
  \citenamefont {Weiss}, \citenamefont {Sch{\"o}nenberger}, \citenamefont
  {Warburton}, \citenamefont {Zbinden},\ and\ \citenamefont
  {Bussi{\`e}res}}]{Cal18apl}%
  \BibitemOpen
  \bibfield  {author} {\bibinfo {author} {\bibfnamefont {M.}~\bibnamefont
  {Caloz}}, \bibinfo {author} {\bibfnamefont {M.}~\bibnamefont {Perrenoud}},
  \bibinfo {author} {\bibfnamefont {C.}~\bibnamefont {Autebert}}, \bibinfo
  {author} {\bibfnamefont {B.}~\bibnamefont {Korzh}}, \bibinfo {author}
  {\bibfnamefont {M.}~\bibnamefont {Weiss}}, \bibinfo {author} {\bibfnamefont
  {Ch.}\ \bibnamefont {Sch{\"o}nenberger}}, \bibinfo {author} {\bibfnamefont
  {R.~J.}\ \bibnamefont {Warburton}}, \bibinfo {author} {\bibfnamefont
  {H.}~\bibnamefont {Zbinden}}, \ and\ \bibinfo {author} {\bibfnamefont
  {F.}~\bibnamefont {Bussi{\`e}res}},\ }\bibfield  {title} {\enquote {\bibinfo
  {title} {High-detection efficiency and low-timing jitter with amorphous
  superconducting nanowire single-photon detectors},}\ }\href {\doibase
  10.1063/1.5010102} {\bibfield  {journal} {\bibinfo  {journal} {Appl. Phys.
  Lett.}\ }\textbf {\bibinfo {volume} {112}},\ \bibinfo {pages} {061103}
  (\bibinfo {year} {2018})}\BibitemShut {NoStop}%
\bibitem [{\citenamefont {Korneeva}\ \emph {et~al.}(2014)\citenamefont
  {Korneeva}, \citenamefont {Mikhailov}, \citenamefont {Pershin}, \citenamefont
  {Manova}, \citenamefont {Divochiy}, \citenamefont {Vakhtomin}, \citenamefont
  {Korneev}, \citenamefont {Smirnov}, \citenamefont {Sivakov}, \citenamefont
  {Devizenko},\ and\ \citenamefont {Goltsman}}]{Kor14sst}%
  \BibitemOpen
  \bibfield  {author} {\bibinfo {author} {\bibfnamefont {Yu.~P.}\ \bibnamefont
  {Korneeva}}, \bibinfo {author} {\bibfnamefont {M.~Yu.}\ \bibnamefont
  {Mikhailov}}, \bibinfo {author} {\bibfnamefont {Yu.~P.}\ \bibnamefont
  {Pershin}}, \bibinfo {author} {\bibfnamefont {N.~N.}\ \bibnamefont {Manova}},
  \bibinfo {author} {\bibfnamefont {A.~V.}\ \bibnamefont {Divochiy}}, \bibinfo
  {author} {\bibfnamefont {Yu.~B.}\ \bibnamefont {Vakhtomin}}, \bibinfo
  {author} {\bibfnamefont {A.~A.}\ \bibnamefont {Korneev}}, \bibinfo {author}
  {\bibfnamefont {K.~V.}\ \bibnamefont {Smirnov}}, \bibinfo {author}
  {\bibfnamefont {A.~G.}\ \bibnamefont {Sivakov}}, \bibinfo {author}
  {\bibfnamefont {A.~Yu.}\ \bibnamefont {Devizenko}}, \ and\ \bibinfo {author}
  {\bibfnamefont {G.~N.}\ \bibnamefont {Goltsman}},\ }\bibfield  {title}
  {\enquote {\bibinfo {title} {Superconducting single-photon detector made of
  {MoSi} film},}\ }\href
  {https://doi.org/10.1088%2F0953-2048%2F27%2F9%2F095012} {\bibfield  {journal}
  {\bibinfo  {journal} {Supercond. Sci. Technol.}\ }\textbf {\bibinfo {volume}
  {27}},\ \bibinfo {pages} {095012} (\bibinfo {year} {2014})}\BibitemShut
  {NoStop}%
\bibitem [{\citenamefont {Dobrovolskiy}\ \emph {et~al.}(2015)\citenamefont
  {Dobrovolskiy}, \citenamefont {Kompaniiets}, \citenamefont {Sachser},
  \citenamefont {Porrati}, \citenamefont {Gspan}, \citenamefont {Plank},\ and\
  \citenamefont {Huth}}]{Dob15bjn}%
  \BibitemOpen
  \bibfield  {author} {\bibinfo {author} {\bibfnamefont {O.~V.}\ \bibnamefont
  {Dobrovolskiy}}, \bibinfo {author} {\bibfnamefont {M.}~\bibnamefont
  {Kompaniiets}}, \bibinfo {author} {\bibfnamefont {R.}~\bibnamefont
  {Sachser}}, \bibinfo {author} {\bibfnamefont {F.}~\bibnamefont {Porrati}},
  \bibinfo {author} {\bibfnamefont {Ch.}\ \bibnamefont {Gspan}}, \bibinfo
  {author} {\bibfnamefont {H.}~\bibnamefont {Plank}}, \ and\ \bibinfo {author}
  {\bibfnamefont {M.}~\bibnamefont {Huth}},\ }\bibfield  {title} {\enquote
  {\bibinfo {title} {Tunable magnetism on the lateral mesoscale by
  post-processing of {Co/Pt} heterostructures},}\ }\href {\doibase
  10.3762/bjnano.6.109} {\bibfield  {journal} {\bibinfo  {journal} {Beilstein
  J. Nanotech.}\ }\textbf {\bibinfo {volume} {6}},\ \bibinfo {pages}
  {1082--1090} (\bibinfo {year} {2015})}\BibitemShut {NoStop}%
\bibitem [{\citenamefont {Porrati}\ \emph {et~al.}(2019)\citenamefont
  {Porrati}, \citenamefont {Barth}, \citenamefont {Sachser}, \citenamefont
  {Dobrovolskiy}, \citenamefont {Seybert}, \citenamefont {Frangakis},\ and\
  \citenamefont {Huth}}]{Por19acs}%
  \BibitemOpen
  \bibfield  {author} {\bibinfo {author} {\bibfnamefont {F.}~\bibnamefont
  {Porrati}}, \bibinfo {author} {\bibfnamefont {S.}~\bibnamefont {Barth}},
  \bibinfo {author} {\bibfnamefont {R.}~\bibnamefont {Sachser}}, \bibinfo
  {author} {\bibfnamefont {O.~V.}\ \bibnamefont {Dobrovolskiy}}, \bibinfo
  {author} {\bibfnamefont {A.}~\bibnamefont {Seybert}}, \bibinfo {author}
  {\bibfnamefont {A.~S.}\ \bibnamefont {Frangakis}}, \ and\ \bibinfo {author}
  {\bibfnamefont {M.}~\bibnamefont {Huth}},\ }\bibfield  {title} {\enquote
  {\bibinfo {title} {Crystalline niobium carbide superconducting nanowires
  prepared by focused ion beam direct writing},}\ }\href {\doibase
  10.1021/acsnano.9b00059} {\bibfield  {journal} {\bibinfo  {journal} {ACS
  Nano}\ }\textbf {\bibinfo {volume} {13}},\ \bibinfo {pages} {6287--6296}
  (\bibinfo {year} {2019})}\BibitemShut {NoStop}%
\bibitem [{\citenamefont {Semenov}\ \emph {et~al.}(2009)\citenamefont
  {Semenov}, \citenamefont {G\"unther}, \citenamefont {B\"ottger},
  \citenamefont {H\"ubers}, \citenamefont {Bartolf}, \citenamefont {Engel},
  \citenamefont {Schilling}, \citenamefont {Ilin}, \citenamefont {Siegel},
  \citenamefont {Schneider}, \citenamefont {Gerthsen},\ and\ \citenamefont
  {Gippius}}]{Sem09prb}%
  \BibitemOpen
  \bibfield  {author} {\bibinfo {author} {\bibfnamefont {A.}~\bibnamefont
  {Semenov}}, \bibinfo {author} {\bibfnamefont {B.}~\bibnamefont {G\"unther}},
  \bibinfo {author} {\bibfnamefont {U.}~\bibnamefont {B\"ottger}}, \bibinfo
  {author} {\bibfnamefont {H.-W.}\ \bibnamefont {H\"ubers}}, \bibinfo {author}
  {\bibfnamefont {H.}~\bibnamefont {Bartolf}}, \bibinfo {author} {\bibfnamefont
  {A.}~\bibnamefont {Engel}}, \bibinfo {author} {\bibfnamefont
  {A.}~\bibnamefont {Schilling}}, \bibinfo {author} {\bibfnamefont
  {K.}~\bibnamefont {Ilin}}, \bibinfo {author} {\bibfnamefont {M.}~\bibnamefont
  {Siegel}}, \bibinfo {author} {\bibfnamefont {R.}~\bibnamefont {Schneider}},
  \bibinfo {author} {\bibfnamefont {D.}~\bibnamefont {Gerthsen}}, \ and\
  \bibinfo {author} {\bibfnamefont {N.~A.}\ \bibnamefont {Gippius}},\
  }\bibfield  {title} {\enquote {\bibinfo {title} {Optical and transport
  properties of ultrathin nbn films and nanostructures},}\ }\href {\doibase
  10.1103/PhysRevB.80.054510} {\bibfield  {journal} {\bibinfo  {journal} {Phys.
  Rev. B}\ }\textbf {\bibinfo {volume} {80}},\ \bibinfo {pages} {054510}
  (\bibinfo {year} {2009})}\BibitemShut {NoStop}%
\bibitem [{\citenamefont {Romijn}\ \emph {et~al.}(1982)\citenamefont {Romijn},
  \citenamefont {Klapwijk}, \citenamefont {Renne},\ and\ \citenamefont
  {Mooij}}]{Rom82prb}%
  \BibitemOpen
  \bibfield  {author} {\bibinfo {author} {\bibfnamefont {J.}~\bibnamefont
  {Romijn}}, \bibinfo {author} {\bibfnamefont {T.~M.}\ \bibnamefont
  {Klapwijk}}, \bibinfo {author} {\bibfnamefont {M.~J.}\ \bibnamefont {Renne}},
  \ and\ \bibinfo {author} {\bibfnamefont {J.~E.}\ \bibnamefont {Mooij}},\
  }\bibfield  {title} {\enquote {\bibinfo {title} {Critical pair-breaking
  current in superconducting aluminum strips far below ${T}_{c}$},}\ }\href
  {\doibase 10.1103/PhysRevB.26.3648} {\bibfield  {journal} {\bibinfo
  {journal} {Phys. Rev. B}\ }\textbf {\bibinfo {volume} {26}},\ \bibinfo
  {pages} {3648--3655} (\bibinfo {year} {1982})}\BibitemShut {NoStop}%
\bibitem [{\citenamefont {Clem}\ and\ \citenamefont {Kogan}(2012)}]{Cle12prb}%
  \BibitemOpen
  \bibfield  {author} {\bibinfo {author} {\bibfnamefont {John~R.}\ \bibnamefont
  {Clem}}\ and\ \bibinfo {author} {\bibfnamefont {V.~G.}\ \bibnamefont
  {Kogan}},\ }\bibfield  {title} {\enquote {\bibinfo {title} {Kinetic impedance
  and depairing in thin and narrow superconducting films},}\ }\href {\doibase
  10.1103/PhysRevB.86.174521} {\bibfield  {journal} {\bibinfo  {journal} {Phys.
  Rev. B}\ }\textbf {\bibinfo {volume} {86}},\ \bibinfo {pages} {174521}
  (\bibinfo {year} {2012})}\BibitemShut {NoStop}%
\bibitem [{\citenamefont {Plourde}\ \emph {et~al.}(2001)\citenamefont
  {Plourde}, \citenamefont {Van~Harlingen}, \citenamefont {Vodolazov},
  \citenamefont {Besseling}, \citenamefont {Hesselberth},\ and\ \citenamefont
  {Kes}}]{Plo01prb}%
  \BibitemOpen
  \bibfield  {author} {\bibinfo {author} {\bibfnamefont {B.~L.~T.}\
  \bibnamefont {Plourde}}, \bibinfo {author} {\bibfnamefont {D.~J.}\
  \bibnamefont {Van~Harlingen}}, \bibinfo {author} {\bibfnamefont {D.~Yu.}\
  \bibnamefont {Vodolazov}}, \bibinfo {author} {\bibfnamefont {R.}~\bibnamefont
  {Besseling}}, \bibinfo {author} {\bibfnamefont {M.~B.~S.}\ \bibnamefont
  {Hesselberth}}, \ and\ \bibinfo {author} {\bibfnamefont {P.~H.}\ \bibnamefont
  {Kes}},\ }\bibfield  {title} {\enquote {\bibinfo {title} {Influence of edge
  barriers on vortex dynamics in thin weak-pinning superconducting strips},}\
  }\href {\doibase 10.1103/PhysRevB.64.014503} {\bibfield  {journal} {\bibinfo
  {journal} {Phys. Rev. B}\ }\textbf {\bibinfo {volume} {64}},\ \bibinfo
  {pages} {014503} (\bibinfo {year} {2001})}\BibitemShut {NoStop}%
\bibitem [{\citenamefont {Ilin}\ \emph {et~al.}(2014)\citenamefont {Ilin},
  \citenamefont {Henrich}, \citenamefont {Luck}, \citenamefont {Liang},
  \citenamefont {Siegel},\ and\ \citenamefont {Vodolazov}}]{Ili14prb}%
  \BibitemOpen
  \bibfield  {author} {\bibinfo {author} {\bibfnamefont {K.}~\bibnamefont
  {Ilin}}, \bibinfo {author} {\bibfnamefont {D.}~\bibnamefont {Henrich}},
  \bibinfo {author} {\bibfnamefont {Y.}~\bibnamefont {Luck}}, \bibinfo {author}
  {\bibfnamefont {Y.}~\bibnamefont {Liang}}, \bibinfo {author} {\bibfnamefont
  {M.}~\bibnamefont {Siegel}}, \ and\ \bibinfo {author} {\bibfnamefont
  {D.~Yu.}\ \bibnamefont {Vodolazov}},\ }\bibfield  {title} {\enquote {\bibinfo
  {title} {Critical current of {Nb}, {NbN}, and {TaN} thin-film bridges with
  and without geometrical nonuniformities in a magnetic field},}\ }\href
  {\doibase 10.1103/PhysRevB.89.184511} {\bibfield  {journal} {\bibinfo
  {journal} {Phys. Rev. B}\ }\textbf {\bibinfo {volume} {89}},\ \bibinfo
  {pages} {184511} (\bibinfo {year} {2014})}\BibitemShut {NoStop}%
\bibitem [{\citenamefont {Maksimova}(1998)}]{Mak98pss}%
  \BibitemOpen
  \bibfield  {author} {\bibinfo {author} {\bibfnamefont {G.~M.}\ \bibnamefont
  {Maksimova}},\ }\bibfield  {title} {\enquote {\bibinfo {title} {Mixed state
  and critical current in narrow semiconducting films},}\ }\href {\doibase
  10.1134/1.1130618} {\bibfield  {journal} {\bibinfo  {journal} {Phys. Sol.
  Stat.}\ }\textbf {\bibinfo {volume} {40}},\ \bibinfo {pages} {1607--1610}
  (\bibinfo {year} {1998})}\BibitemShut {NoStop}%
\bibitem [{\citenamefont {Doettinger}\ \emph {et~al.}(1995)\citenamefont
  {Doettinger}, \citenamefont {Huebener},\ and\ \citenamefont
  {K\"uhle}}]{Doe95pcs}%
  \BibitemOpen
  \bibfield  {author} {\bibinfo {author} {\bibfnamefont {S.G.}\ \bibnamefont
  {Doettinger}}, \bibinfo {author} {\bibfnamefont {R.P.}\ \bibnamefont
  {Huebener}}, \ and\ \bibinfo {author} {\bibfnamefont {A.}~\bibnamefont
  {K\"uhle}},\ }\bibfield  {title} {\enquote {\bibinfo {title} {Electronic
  instability during vortex motion in cuprate superconductors regime of low and
  high magnetic fields},}\ }\href {\doibase
  http://dx.doi.org/10.1016/0921-4534(95)00411-4} {\bibfield  {journal}
  {\bibinfo  {journal} {Physica C}\ }\textbf {\bibinfo {volume} {251}},\
  \bibinfo {pages} {285 -- 289} (\bibinfo {year} {1995})}\BibitemShut {NoStop}%
\bibitem [{\citenamefont {Babic}\ \emph {et~al.}(2004)\citenamefont {Babic},
  \citenamefont {Bentner}, \citenamefont {S\"urgers},\ and\ \citenamefont
  {Strunk}}]{Bab04prb}%
  \BibitemOpen
  \bibfield  {author} {\bibinfo {author} {\bibfnamefont {D.}~\bibnamefont
  {Babic}}, \bibinfo {author} {\bibfnamefont {J.}~\bibnamefont {Bentner}},
  \bibinfo {author} {\bibfnamefont {C.}~\bibnamefont {S\"urgers}}, \ and\
  \bibinfo {author} {\bibfnamefont {C.}~\bibnamefont {Strunk}},\ }\bibfield
  {title} {\enquote {\bibinfo {title} {Flux-flow instabilities in amorphous
  {${\mathrm{Nb}}_{0.7}{\mathrm{Ge}}_{0.3}$ } microbridges},}\ }\href {\doibase
  10.1103/PhysRevB.69.092510} {\bibfield  {journal} {\bibinfo  {journal} {Phys.
  Rev. B}\ }\textbf {\bibinfo {volume} {69}},\ \bibinfo {pages} {092510--1--4}
  (\bibinfo {year} {2004})}\BibitemShut {NoStop}%
\bibitem [{\citenamefont {Sidorova}\ \emph {et~al.}(2018)\citenamefont
  {Sidorova}, \citenamefont {Kozorezov}, \citenamefont {Semenov}, \citenamefont
  {Korneeva}, \citenamefont {Mikhailov}, \citenamefont {Devizenko},
  \citenamefont {Korneev}, \citenamefont {Chulkova},\ and\ \citenamefont
  {Goltsman}}]{Sid18prb}%
  \BibitemOpen
  \bibfield  {author} {\bibinfo {author} {\bibfnamefont {Mariia~V.}\
  \bibnamefont {Sidorova}}, \bibinfo {author} {\bibfnamefont {A.~G.}\
  \bibnamefont {Kozorezov}}, \bibinfo {author} {\bibfnamefont {A.~V.}\
  \bibnamefont {Semenov}}, \bibinfo {author} {\bibfnamefont {Yu.~P.}\
  \bibnamefont {Korneeva}}, \bibinfo {author} {\bibfnamefont {M.~Yu.}\
  \bibnamefont {Mikhailov}}, \bibinfo {author} {\bibfnamefont {A.~Yu.}\
  \bibnamefont {Devizenko}}, \bibinfo {author} {\bibfnamefont {A.~A.}\
  \bibnamefont {Korneev}}, \bibinfo {author} {\bibfnamefont {G.~M.}\
  \bibnamefont {Chulkova}}, \ and\ \bibinfo {author} {\bibfnamefont {G.~N.}\
  \bibnamefont {Goltsman}},\ }\bibfield  {title} {\enquote {\bibinfo {title}
  {Nonbolometric bottleneck in electron-phonon relaxation in ultrathin {WSi}
  films},}\ }\href {\doibase 10.1103/PhysRevB.97.184512} {\bibfield  {journal}
  {\bibinfo  {journal} {Phys. Rev. B}\ }\textbf {\bibinfo {volume} {97}},\
  \bibinfo {pages} {184512} (\bibinfo {year} {2018})}\BibitemShut {NoStop}%
\bibitem [{\citenamefont {Sidorova}\ \emph {et~al.}(2020)\citenamefont
  {Sidorova}, \citenamefont {Semenov}, \citenamefont {H\"ubers}, \citenamefont
  {Ilin}, \citenamefont {Siegel}, \citenamefont {Charaev}, \citenamefont
  {Moshkova}, \citenamefont {Kaurova}, \citenamefont {Goltsman}, \citenamefont
  {Zhang},\ and\ \citenamefont {Schilling}}]{Sid20prb}%
  \BibitemOpen
  \bibfield  {author} {\bibinfo {author} {\bibfnamefont {M.}~\bibnamefont
  {Sidorova}}, \bibinfo {author} {\bibfnamefont {A.}~\bibnamefont {Semenov}},
  \bibinfo {author} {\bibfnamefont {H.-W.}\ \bibnamefont {H\"ubers}}, \bibinfo
  {author} {\bibfnamefont {K.}~\bibnamefont {Ilin}}, \bibinfo {author}
  {\bibfnamefont {M.}~\bibnamefont {Siegel}}, \bibinfo {author} {\bibfnamefont
  {I.}~\bibnamefont {Charaev}}, \bibinfo {author} {\bibfnamefont
  {M.}~\bibnamefont {Moshkova}}, \bibinfo {author} {\bibfnamefont
  {N.}~\bibnamefont {Kaurova}}, \bibinfo {author} {\bibfnamefont {G.~N.}\
  \bibnamefont {Goltsman}}, \bibinfo {author} {\bibfnamefont {X.}~\bibnamefont
  {Zhang}}, \ and\ \bibinfo {author} {\bibfnamefont {A.}~\bibnamefont
  {Schilling}},\ }\bibfield  {title} {\enquote {\bibinfo {title} {Electron
  energy relaxation in disordered superconducting {NbN} films},}\ }\href
  {\doibase 10.1103/PhysRevB.102.054501} {\bibfield  {journal} {\bibinfo
  {journal} {Phys. Rev. B}\ }\textbf {\bibinfo {volume} {102}},\ \bibinfo
  {pages} {054501} (\bibinfo {year} {2020})}\BibitemShut {NoStop}%
\bibitem [{\citenamefont {Silhanek}\ \emph {et~al.}(2010)\citenamefont
  {Silhanek}, \citenamefont {Milo\ifmmode \check{s}\else
  \v{s}\fi{}evi\ifmmode~\acute{c}\else \'{c}\fi{}}, \citenamefont {Kramer},
  \citenamefont {Berdiyorov}, \citenamefont {Van~de Vondel}, \citenamefont
  {Luccas}, \citenamefont {Puig}, \citenamefont {Peeters},\ and\ \citenamefont
  {Moshchalkov}}]{Sil10prl}%
  \BibitemOpen
  \bibfield  {author} {\bibinfo {author} {\bibfnamefont {A.~V.}\ \bibnamefont
  {Silhanek}}, \bibinfo {author} {\bibfnamefont {M.~V.}\ \bibnamefont
  {Milo\ifmmode \check{s}\else \v{s}\fi{}evi\ifmmode~\acute{c}\else
  \'{c}\fi{}}}, \bibinfo {author} {\bibfnamefont {R.~B.~G.}\ \bibnamefont
  {Kramer}}, \bibinfo {author} {\bibfnamefont {G.~R.}\ \bibnamefont
  {Berdiyorov}}, \bibinfo {author} {\bibfnamefont {J.}~\bibnamefont {Van~de
  Vondel}}, \bibinfo {author} {\bibfnamefont {R.~F.}\ \bibnamefont {Luccas}},
  \bibinfo {author} {\bibfnamefont {T.}~\bibnamefont {Puig}}, \bibinfo {author}
  {\bibfnamefont {F.~M.}\ \bibnamefont {Peeters}}, \ and\ \bibinfo {author}
  {\bibfnamefont {V.~V.}\ \bibnamefont {Moshchalkov}},\ }\bibfield  {title}
  {\enquote {\bibinfo {title} {Formation of stripelike flux patterns obtained
  by freezing kinematic vortices in a superconducting {Pb} film},}\ }\href
  {\doibase 10.1103/PhysRevLett.104.017001} {\bibfield  {journal} {\bibinfo
  {journal} {Phys. Rev. Lett.}\ }\textbf {\bibinfo {volume} {104}},\ \bibinfo
  {pages} {017001} (\bibinfo {year} {2010})}\BibitemShut {NoStop}%
\bibitem [{\citenamefont {Dobrovolskiy}\ \emph
  {et~al.}(2019{\natexlab{b}})\citenamefont {Dobrovolskiy}, \citenamefont
  {Sachser}, \citenamefont {Br{\"a}cher}, \citenamefont {B{\"o}ttcher},
  \citenamefont {Kruglyak}, \citenamefont {Vovk}, \citenamefont {Shklovskij},
  \citenamefont {Huth}, \citenamefont {Hillebrands},\ and\ \citenamefont
  {Chumak}}]{Dob19nph}%
  \BibitemOpen
  \bibfield  {author} {\bibinfo {author} {\bibfnamefont {O.~V.}\ \bibnamefont
  {Dobrovolskiy}}, \bibinfo {author} {\bibfnamefont {R.}~\bibnamefont
  {Sachser}}, \bibinfo {author} {\bibfnamefont {T.}~\bibnamefont
  {Br{\"a}cher}}, \bibinfo {author} {\bibfnamefont {T.}~\bibnamefont
  {B{\"o}ttcher}}, \bibinfo {author} {\bibfnamefont {V.~V.}\ \bibnamefont
  {Kruglyak}}, \bibinfo {author} {\bibfnamefont {R.~V.}\ \bibnamefont {Vovk}},
  \bibinfo {author} {\bibfnamefont {V.~A.}\ \bibnamefont {Shklovskij}},
  \bibinfo {author} {\bibfnamefont {M.}~\bibnamefont {Huth}}, \bibinfo {author}
  {\bibfnamefont {B.}~\bibnamefont {Hillebrands}}, \ and\ \bibinfo {author}
  {\bibfnamefont {A.~V.}\ \bibnamefont {Chumak}},\ }\bibfield  {title}
  {\enquote {\bibinfo {title} {Magnon-fluxon interaction in a
  ferromagnet/superconductor heterostructure},}\ }\href {\doibase
  10.1038/s41567-019-0428-5} {\bibfield  {journal} {\bibinfo  {journal} {Nat.
  Phys.}\ }\textbf {\bibinfo {volume} {15}},\ \bibinfo {pages} {477} (\bibinfo
  {year} {2019}{\natexlab{b}})}\BibitemShut {NoStop}%
\end{thebibliography}
%

\end{document}